\documentclass[11pt,a4paper]{article}
\pdfoutput=1
\usepackage{jheppub}
\usepackage{rotating}

\usepackage{wrapfig}

\usepackage{amsfonts}
\usepackage{amsmath}
\usepackage{slashed}
\usepackage{amssymb}
\usepackage{latexsym}
\usepackage{multirow}
\usepackage{hyperref}
\usepackage{enumitem}
\hypersetup{colorlinks=true,citecolor=red,linkcolor=magenta,urlcolor=blue}
\usepackage{relsize}
\usepackage{booktabs}
\usepackage{subfigure}
\usepackage{cleveref}
\usepackage{fancyhdr}
\usepackage{float}
\usepackage{graphicx}
\usepackage{microtype}
\usepackage{tabularx}
\usepackage{xcolor}
\usepackage{siunitx,adjustbox,bm}
\usepackage[bottom]{footmisc}
\usepackage{makecell}
\usepackage[normalem]{ulem}

\newcommand{\mathematicanb}{\href{https://github.com/shakeel-hep/D8-WCs}{\underline{Mathematica notebook}}}

\newcommand{\sr}[1]{{\textcolor{black}{#1}}}

\newcommand{\nn}{\nonumber}

\begin{document}
\allowdisplaybreaks
\title{Integrating out heavy scalars with modified EOMs: matching computation of dimension-eight SMEFT coefficients}
\abstract{The shift in focus towards searches for physics beyond the Standard Model (SM) employing model-independent Effective Field Theory (EFT) methods necessitates a rigorous approach to matching to guarantee the validity of the obtained results and constraints. 
The limits on the leading dimension-six EFT effects can be rather inaccurate for LHC searches that suffer from large uncertainties while exploring an extensive energy range. Similarly, precise measurements can, in principle, test the subleading effects of the operator expansion.
In this work, we present an algorithmic approach to automatise matching computations for dimension-eight operators for generic scalar extensions with proper implementation of equations of motion. 
We devise a step-by-step procedure to obtain the dimension-eight Wilson coefficients (WCs) in a non-redundant basis to arrive at complete matching results.
We apply this formalism to a range of scalar extensions of the SM and provide tree-level and loop-suppressed results.  
Finally, we discuss the relevance of the dimension-eight operators for a range of phenomenological analyses, particularly focusing on Higgs and electroweak physics.}	
\author[1]{Upalaparna Banerjee, Joydeep Chakrabortty,}
\author[2]{Christoph Englert,}
\author[1]{Shakeel Ur Rahaman,}
\author[3]{and Michael Spannowsky}
\emailAdd{upalab@iitk.ac.in, joydeep@iitk.ac.in, christoph.englert@glasgow.ac.uk, shakel@iitk.ac.in, michael.spannowsky@durham.ac.uk}
\affiliation[1]{Indian Institute of Technology Kanpur, Kalyanpur, Kanpur 208016, Uttar Pradesh, India}
\affiliation[2]{School of Physics \& Astronomy, University of Glasgow, Glasgow G12 8QQ, U.K.}
\affiliation[3]{Institute for Particle Physics Phenomenology, Department of Physics, Durham University, Durham DH1 3LE, U.K.}
	
\preprint{}

\maketitle
	
\section{Introduction}
Searches for physics beyond the Standard Model (BSM) chiefly performed at the Large Hadron Collider (LHC) have, so far, not revealed any significant deviation from the Standard Model (SM) predictions. This is puzzling, on the one hand, given the SM's plethora of known flaws and shortcomings. On the other hand, these findings have motivated the application of model-independent techniques employing Effective Field Theory (EFT)~\cite{Weinberg:1978kz} to LHC data. The EFT approach breaks away from the assumption of concrete model-dependent correlations, thus opening up the possibility of revealing new (and perhaps non-canonical) BSM interactions through a holistic approach to data correlation interpretation. The inherent assumption of such an approach is that there is a significant mass gap between the BSM spectrum and the (process-dependent) characteristic energy scale at which the LHC operates to ``integrate out'' BSM states to obtain a low energy effective description that is determined by the SM's particle and symmetry content.

Efforts to apply EFT to the multi-scale processes of the LHC environment have received considerable interest recently, reaching from theory-led proof-of-principle fits to LHC data~\cite{Buckley:2015lku,Englert:2015hrx,Butter:2016cvz,Brivio:2019ius,Ellis:2020unq,Ethier:2021bye} (with a history of almost a decade) over the adoption of these techniques by the LHC experiments~(e.g.~\cite{CMS:2020lrr,ATLAS:2022xyx} for recent examples), to perturbative improvements of the formalism~\cite{Alonso:2013hga,Jenkins:2013wua,Jenkins:2013sda,Jenkins:2013zja,Hartmann:2015aia,Dedes:2017zog,Dedes:2018seb,Dawson:2018liq,Dawson:2018pyl,Dawson:2019clf,Dawson:2021ofa,Machado:2022ozb}. In doing so, most attention has been devoted to SMEFT at dimension-six level~\cite{Grzadkowski:2010es}
\begin{equation}
\label{eq:d6}
{\cal{L}} = {\cal{L}}_{\text{SM}} + \sum_i {c_i\over \Lambda^2} {\cal{O}}_i\,.
\end{equation}

While EFT is a formidable tool to put correlations at the forefront of BSM searches, the significant energy coverage of the LHC can lead to blurred sensitivity estimates even in instances when Eq.~\eqref{eq:d6} is a sufficiently accurate expansion. When pushing the cut-off scale $\Lambda$ to large values, the experimental sensitivity to deviations from the SM can be too small to yield perturbatively meaningful or relevant constraints when matched to concrete UV scenarios (see e.g.~\cite{Englert:2019rga}). In contrast, dimension-eight contributions can be sizeable when the new physics cut-off $\Lambda$ is comparably low in the case of more significant BSM signals at the LHC. To understand the ramifications for concrete UV models, it is then important to (i) have a flexible approach to mapping out the dimension-eight interactions and (ii) gauge the importance of dimension-eight contributions relative to those of dimension-six to quantitatively assess the error of the (potential) dimension-eight truncation.

A common bottleneck in constructing EFT interactions is removing redundancies. This is historically evidenced by the emergence of the so-called Warsaw basis~\cite{Grzadkowski:2010es}. Equations of motion are typically considered in eliminating redundant operators. Still, they are not identical to general (non-linear) field redefinitions, which are the actual redundant parameters of the field theory~\cite{Georgi:1991ch,Manohar:1997qy,Manohar:1996cq,Einhorn:2001kj,Barzinji:2018xvu}. When truncating a given operator dimension, this can be viewed as a scheme-dependence not too unfamiliar from renormalisable theories, however, with less controlled side-effects when the new physics scale is comparably low. Additional operator structures need to be included to elevate classical equations of motion to field re-definitions~\cite{Barzinji:2018xvu,Criado:2018sdb} to achieve a consistent classification at the dimension-eight level.

In this work, we devise a generic approach to this issue which enables us to provide a complete framework to match any dimension-eight structure that emerges in the process of integrating-out a heavy non-SM scalar and obtaining the form of the WCs. Along the way of systematically re-organising the operators into a non-redundant basis, resembling the one discussed in Refs.~\cite{Murphy:2020rsh,Li:2020gnx}, we show that removing the higher-derivative operators produced at the dimension-six level itself can induce a non-negligible effect on dimension-eight matching coefficients along with the direct contribution to the same which can be computed following the familiar methodologies of the Covariant Derivative Expansion (CDE)~\cite{Henning:2014wua,Henning:2016lyp} of the path integral~\cite{Gaillard:1985uh,Cheyette:1987qz,Dittmaier:2021fls}, or diagrammatic approach~\cite{Zhang:2016pja,Jiang:2018pbd}. Finally, it is worth mentioning that, even though the one-loop effective action at dimension eight is yet to be formulated, it is possible to receive equally suppressed, loop-induced corrections from the dimension-six coefficients computed precisely at one-loop. These can present themselves as the leading order contributions for the WCs, which generally appear at one-loop.

This paper is organised as follows: in Sec.~\ref{sec:complete-tree-Lag}, we discuss the implementation of the Higgs field equation of motion and study its equivalence with field redefinitions. 
This gives rise to the desired dimension-eight operator structures after removing redundancies (Sec.~\ref{subsec:dim-8-tree-Lag}).
Our approach is tested and validated against available results for the real triplet scalar extension in Sec.~\ref{sec:x-validation}. In Sec.~\ref{sec:example}, the matching coefficients are presented explicitly considering a range of scalar extensions of the SM. Finally, the significance of the dimension-eight operators is analysed based on observables in a model-dependent manner in Sec.~\ref{sec:classification}. We conclude in Sec.~\ref{sec:conc}.

\section{Complete matching at dimension-eight}
\label{sec:complete-tree-Lag}
We start by studying the structures of the higher-dimensional operators that can arise from heavy scalar extensions of the SM generically once the heavy field $(\Phi)$ is integrated out. The most generalised structure of the renormalised Lagrangian involving heavy scalars can be written as~\cite{Henning:2014wua,Brivio:2017vri}:
\begin{equation}
\label{eq:General-Lagrangian}
	\mathcal{L}\,\big[\Phi\big] \supset \Phi^{\dagger}(P^2-m^2-U(x))\,\Phi+(\Phi^{\dagger}\,B(x)+\text{h.c.})+\frac{1}{4}\,\lambda_{\Phi}\,(\Phi^{\dagger}\Phi)^2.
\end{equation}	
Here, $U(x)$ and $B(x)$ contain the interactions that are quadratic and linear in $\Phi$, respectively, and only involve the lighter degrees of freedom. Once $\Phi$ is integrated out, we obtain a tower of operators that can be arranged according to their canonical dimension. It is important to note that the operators generated in this process might not be independent. Depending on phenomenological considerations, several sets of operators are defined in the literature. A set of dimension-six operators was prescribed in Ref.~\cite{BUCHMULLER1986621}. It was improved by systematically removing the redundant structures and promoting it to form a complete non-redundant basis in Ref.~\cite{Grzadkowski:2010es}\footnote{A minimal set of four-fermionic operators is also constructed in Ref.~\cite{Aguilar-Saavedra:2010uur}.}, popularly known as the \emph{Warsaw basis}. There is another set of operators known as the \emph{Green's set}~\cite{Gherardi:2020det,Carmona:2021xtq,Chala:2021cgt}, which is over-complete. The operators here are independent under the Fierz identities and integration by parts but otherwise redundant on account of equations of motion\footnote{Here we are being ambiguous about the use of the equation of motion or field redefinition in removing the redundancies, see Ref.~\cite{Georgi:1991ch, Criado:2018sdb} for more details.}. This source of redundancy contributes to higher dimensional operators. In this paper, we  use the Mathematica package \texttt{CoDEx}~\cite{Bakshi:2018ics} to generate WCs of the operators in the \emph{SILH set}~\cite{Giudice:2007fh,Contino:2013kra} up to one-loop, including the relevant redundant terms. 
Since we are interested in the corrections to the dimension-eight coefficients resulting from the dimension-six redundant structures, we recast the \emph{SILH} operators into \emph{Green's set}-like structures\footnote{We call it ``\emph{Green's set}-like structures" because we differ in some redundant operator structures as defined in Ref.~\cite{Gherardi:2020det}, see appendix \ref{app:op-structures} for more details.} to single out redundant and non-redundant operators using the following equations:
  	  \begin{eqnarray}\label{eq:SILH_to_mixed}
		\mathcal{Q}_{H}&\equiv&\frac{1}{2}\partial_{\mu}(H^{\dagger}H)\partial^{\mu}(H^{\dagger}H) = -\frac{1}{2}(H^{\dagger}H)\square(H^{\dagger}H),\nonumber\\
		\mathcal{Q}_{T}&\equiv&\frac{1}{2}\big[H^{\dagger}\overleftrightarrow{\mathcal{D}}^{\mu}H\big]\big[H^{\dagger}\overleftrightarrow{\mathcal{D}}^{\mu}H\big]= -2\,(\mathcal{D}_{\mu}H^{\dagger}H)(H^{\dagger}\mathcal{D}_{\mu}H)-\frac{1}{2}(H^{\dagger}H)\square(H^{\dagger}H),\nonumber\\
		\mathcal{Q}_{R}&\equiv&(H^{\dagger}H)(\mathcal{D}_{\mu}H^{\dagger}D^{\mu}H) = \frac{1}{2}\big[(H^{\dagger}H)\square(H^{\dagger}H)-\boldsymbol{(H^{\dagger}H)(\mathcal{D}^2 H^{\dagger}H+H^{\dagger}\mathcal{D}^2 H)}\big],\nonumber\\
		\mathcal{Q}_{\mathcal{D}}&\equiv&\mathcal{D}^2H^{\dagger}D^2H = \boldsymbol{-\frac{1}{2}(Y_{pq}\,(\overline{\psi}_{p}\psi_q)\,\mathcal{D}^2H\,+\,\textbf{h.c.})}-\boldsymbol{\lambda'\,(H^{\dagger}H)(\mathcal{D}^2 H^{\dagger}H+H^{\dagger}\mathcal{D}^2 H)},\nonumber\\
		\mathcal{Q}_{2W}&\equiv&-\frac{1}{2} \big(\mathcal{D}_{\mu}W^I_{\mu\nu}\big)^2=-\boldsymbol{\frac{g_{_W}^2}{32}Y_{pq}^{-1}\big((\overline{\psi}_p\,\psi_q)\,\mathcal{D}^2H\,+\,\textbf{h.c.}\big)}-\frac{ig_{_W}^2}{4}\big(\overline{\psi}_p\gamma^{\mu}\tau^I\psi_p\big)(H^{\dagger}i\overleftrightarrow{\mathcal{D}_{\mu}^I}H) \nonumber \\ & & \hspace{0cm} +\frac{g_{_W}^2}{4}\lambda'^{2}Y_{pq}^{-1}Y_{qp}^{-1}(H^{\dagger}H)^3+\frac{g_{_W}^2}{8}(\mathcal{D}^{
		\mu}H^{\dagger}H)(H^{\dagger}\mathcal{D}_{\mu}H)-\frac{g_{_W}^2}{16}\mathcal{Q}_{R} \nonumber \\ && \hspace{0cm}+\boldsymbol{\frac{g_{_W}^2}{32}(1+2\lambda'Y_{pq}^{-1}Y_{qp}^{-1})\,(H^{\dagger}H)(\mathcal{D}^2 H^{\dagger}H+H^{\dagger}\mathcal{D}^2 H)},\nonumber\\
		\mathcal{Q}_{2B}&\equiv&-\frac{1}{2} \big(\partial_{\mu}B_{\mu\nu}\big)^2=\boldsymbol{-\frac{g_{_Y}^{2}}{4}Y_{pq}^{-1}\big((\overline{\psi}_p\,\psi_q)\,\mathcal{D}^2H\,+\,\textbf{h.c.}\big)}-ig_{_Y}^{2}\big(\overline{\psi}_{p}\gamma^{\mu}\psi_p\big)(H^{\dagger}i\overleftrightarrow{\mathcal{D}_{\mu}}H)\nonumber\\
		&&\hspace{0cm}+g_{_Y}^{2}\,\mathcal{Q}_{R}+2g_{_Y}^{2}\lambda'^{2}\,Y_{pq}^{-1}\,Y_{qp}^{-1}(H^{\dagger}H)^3+\boldsymbol{\,\frac{g_{_Y}^{2}}{2}(1+\lambda'Y_{pq}^{-1}Y_{qp}^{-1})\,(H^{\dagger}H)(\mathcal{D}^2 H^{\dagger}H+H^{\dagger}\mathcal{D}^2 H)},\nonumber\\
		\mathcal{Q}_{2G}&\equiv&-\frac{1}{2} \big(\mathcal{D}_{\mu}G^a_{\mu\nu}\big)^2=\boldsymbol{-\frac{\,g_{_G}^2}{3}Y_{pq}^{-1}\big((\overline{\psi}_p\,\psi_q)\,\mathcal{D}^2H\,+\,\textbf{h.c.}\big)}+\frac{4g_{_G}^2}{3}\lambda^{'2}Y_{pq}^{-1}Y_{qp}^{-1}(H^{\dagger}H)^3\nonumber\\
		&&\boldsymbol{+\frac{2g_{G}^2}{3}\lambda^{'}Y_{pq}^{-1}Y_{qp}^{-1}\,(H^{\dagger}H)(\mathcal{D}^2 H^{\dagger}H+H^{\dagger}\mathcal{D}^2 H)},\nonumber\\
		\mathcal{Q}_{W}&\equiv&ig_{_W}(H^{\dagger}\tau^I\overleftrightarrow{\mathcal{D}}^{\mu}H)\mathcal{D}^{\nu}W^I_{\mu\nu}=\frac{ig_{_W}^2}{2}(H^{\dagger}i\overleftrightarrow{\mathcal{D}}^{I}_{\mu}H)(\overline{\psi}\gamma^{\mu}\tau^I\psi)-\frac{g_{_W}^2}{8}(\mathcal{D}_{\mu}H^{\dagger}H)(H^{\dagger
		}\mathcal{D}^{\mu}H)\nonumber\\
		&&\hspace{0cm}+\frac{g_{_W}^2}{16}\,\mathcal{Q}_{R}-\boldsymbol{\frac{g_{_W}^2}{32}(H^{\dagger}H)(\mathcal{D}^2 H^{\dagger}H+H^{\dagger}\mathcal{D}^2 H)},\nonumber\\
		\mathcal{Q}_{B}&\equiv&ig_{_Y}(H^{\dagger}\overleftrightarrow{\mathcal{D}}^{\mu}H)\partial^{\nu}B_{\mu\nu}=ig_{_Y}^{2}(H^{\dagger}\overleftrightarrow{\mathcal{D}}^{\mu}H)(\overline{\psi}\gamma^{\mu}\psi)-2g_{_Y}^{2}\mathcal{Q}_R \nonumber \\ & & \hspace{0cm}-\boldsymbol{g_{_Y}^{2}(H^{\dagger}H)(\mathcal{D}^2 H^{\dagger}H+H^{\dagger}\mathcal{D}^2 H)},\nonumber\\	\mathcal{Q}_{WW}&\equiv&g_{_W}^2(H^{\dagger}H)W^I_{\mu\nu}W^{I,\mu\nu},\nonumber\\
		\mathcal{Q}_{BB}&\equiv&g_{_Y}^{2}(H^{\dagger}H)B_{\mu\nu}B^{\mu\nu},\nonumber\\
		\mathcal{Q}_{WB}&\equiv&2g_{_W}g_{_Y}(H^{\dagger}\tau^IH)W^I_{\mu\nu}B^{\mu\nu},\nonumber\\
		\mathcal{Q}_{GG}&\equiv&g_{_G}^{2}(H^{\dagger}H)G^a_{\mu\nu}G^{a\,\mu\nu},	
	\end{eqnarray}
where $Y_{pq}$ denotes the SM Yukawa coupling matrix, $\{p,\,q\} \in (1,2,3)$ are the flavour indices. We denote the Wilson coefficients of the SILH operators as $\mathcal{C}_i$ with $i$ labelling the operators in Eqs.~\eqref{eq:SILH_to_mixed}.
Taking into account all the $H$-involved structures that can appear from a scalar extension of the SM, one can write: 
\begin{multline}\label{eq:Lagrangian}
   	\mathcal{L} = \mathcal{L}_{\text{SM}}^{(4)}+\widetilde{\lambda}\,(H^{\dagger}H)^2+\zeta^{(6)}_{1}\,(H^{\dagger}H)^3+\zeta^{(6)}_{2}\,(H^{\dagger}H)\square(H^{\dagger}H)
	+\zeta^{(6)}_3(\mathcal{D}_{\mu}H^{\dagger}H)(H^{\dagger}\mathcal{D}_{\mu}H)\\
    	+\zeta^{(6)}_4 (H^{\dagger}H)(B_{\mu\nu}B^{\mu\nu})
	+\zeta^{(6)}_5(H^{\dagger}H)(W_{\mu\nu}^{I}W^{I \mu\nu})
	+\zeta^{(6)}_6(H^{\dagger}\tau^{I}H)(B_{\mu\nu}W^{I\mu\nu}) \\
	+\zeta^{(6)}_7(H^{\dagger}H)(G^{a}_{\mu\nu}G^{a\mu\nu})
	+ \zeta^{(6)}_{8,1} (H^{\dagger}i\overleftrightarrow{\mathcal{D}}_{\mu}H)(\overline{\psi}\,\gamma^{\mu}\,\psi) 
	+ \zeta^{(6)}_{8,2} (H^{\dagger}i\overleftrightarrow{\mathcal{D}}_{\mu}^{I}H)(\overline{\psi}\,\tau^{I}\gamma^{\mu}\,\psi)\\
    	+\,\boldsymbol{\xi^{(6)}_{1}(H^{\dagger}H)(\mathcal{D}^2 H^{\dagger}H+H^{\dagger}\mathcal{D}^2 H)}
	+\,\boldsymbol{\xi^{(6)}_2\left[\,(\overline{\psi}\,\psi)\,\mathcal{D}^2H\,+\,\textbf{h.c.}\right]}.
\end{multline}   

We highlight the redundant terms in Eq.~\eqref{eq:Lagrangian} in bold font; they need to be removed. The coefficients of the \emph{Green's set}-like structures in Eq.~\eqref{eq:Lagrangian} can be expressed in terms of SILH coefficients through the relations given in Table~\ref{Tab:SILH_to_mixed}.

\begin{table}[h]
	\centering \small
	\renewcommand{\arraystretch}{2.1}
	\begin{tabular}{|c|c|}
		\hline
		\makecell{Coefficients of \\ \emph{Green's set}-like \\operators}&
		Relation in terms of SILH coefficients\\
		
		\hline

		$\zeta_{1}^{(6)}$&
		$\big[\mathcal{C}_6+\frac{g_{_W}^2}{4}\lambda^{'2}Y_{pq}^{-1}Y_{qp}^{-1}\mathcal{C}_{2W}+2g_{_Y}^{2}\lambda^{'2}Y_{pq}^{-1}Y_{qp}^{-1}\mathcal{C}_{2B}+\frac{4g_{G}^2}{3}\lambda^{'2}Y_{pq}^{-1}Y_{qp}^{-1}\mathcal{C}_{2G}\big]$\\
		\hline		
		
		$\zeta_{2}^{(6)}$&
		$\big[-\frac{1}{2}\mathcal{C}_H-\frac{1}{2}\mathcal{C}_{T}+\frac{1}{2}\mathcal{C}_{R}-\frac{g_{_W}^2}{32}\mathcal{C}_{2W}+\frac{g_{_Y}^{2}}{2}\mathcal{C}_{2B}+\frac{g_{_W}^2}{32}\mathcal{C}_{W}-g_{_Y}^{2}\mathcal{C}_B\big]$\\
		\hline
		
		$\zeta_3^{(6)}$&
		$\big[-2\mathcal{C}_T+\frac{g_{_W}^2}{8}\mathcal{C}_{2W}-\frac{g_{_W}^2}{8}\mathcal{C}_{W}\big]$\\
		\hline
		
		$\zeta_{4}^{(6)}$&
		$g_{_Y}^{2}\,\mathcal{C}_{BB}$\\
		\hline
		
		$\zeta_5^{(6)}$&
		$g_{_W}^{2}\,\mathcal{C}_{WW}$\\
		\hline
		
		$\zeta_6^{(6)}$&
		$2g_{_W}g_{_Y}\,\mathcal{C}_{WB}$\\
		\hline
		
		$\zeta_7^{(6)}$&
		$g{_G}^2\,\mathcal{C}_{GG}$\\
		\hline		
		
		$\zeta_{8,1}^{(6)}$&
		$\big[-ig_{_Y}^{2}\,\mathcal{C}_{2B}+ig_{_Y}^{2}\,\mathcal{C}_{B}\big]$\\
		\hline
		
		$\zeta_{8,2}^{(6)}$&
		$\big[-\frac{ig_{_W}^2}{4}\,\mathcal{C}_{2W}+\frac{ig_{_W}^{2}}{2}\,\mathcal{C}_{W}\big]$\\
		\hline  		
		
		\multirow{2}{*}{$\xi_{1}^{(6)}$}&
		$\big[-\frac{1}{2}\mathcal{C}_R+\frac{g_{_W}^2}{32}(1+2\lambda^{'}Y_{pq}^{-1}Y_{qp}^{-1})\mathcal{C}_{2W}+\frac{g_{_Y}^{2}}{2}(1+\lambda^{'}Y_{pq}^{-1}Y_{qp}^{-1})\mathcal{C}_{2B}$\\
		
		&$+\frac{2g_{G}^2}{3}\lambda^{'}Y_{pq}^{-1}Y_{qp}^{-1}\,\mathcal{C}_{2G}-\frac{g_{_W}^2}{16}\mathcal{C}_{W}\big]-\lambda^{'}\mathcal{C}_{\mathcal{D}}$\\
		\hline
		
	    $\xi_{2}^{(6)}$&
		$-Y_{pq}^{-1}\big[\frac{g_{_W}^2}{32}\,\mathcal{C}_{2W}+\frac{g_{_Y}^{2}}{4}\,\mathcal{C}_{2B}+\frac{\,g_{_G}^2}{3}\,\mathcal{C}_{2G}\big]-\frac{1}{2}Y_{pq}\,\mathcal{C}_{\mathcal{D}}$\\
		\hline

	\end{tabular}
	\caption{\small  Translation of \emph{SILH} coefficients into the \emph{Green's set}-like form. Here, $Y_{pq}$ denotes the SM Yuakawa coupling, $\{p,\,q\} \in (1,2,3)$ are the flavour indices.}
	\label{Tab:SILH_to_mixed}
\end{table}
We now discuss our approach to properly implement the Higgs equation of motion (EOM) to compute the corrections to dimension-eight coefficients.

\subsection{Removing redundancies: Field Redefinition \& Higgs Field EOM}
\label{subsec:EOM_vs_field_redefinition}
In QFT, the experimentally observable quantities are related to the S-matrix elements, which remain invariant under field redefinition. Naively, this can be inferred from the fact that when calculating correlation functions using the path integral formalism, the field is just an integration variable. The correlation functions and S-matrix elements can be connected by the LSZ-reduction formula~\cite{Lehmann:1954rq,Manohar:2018aog}. In the case of a renormalisable Lagrangian, we exploit this freedom and rewrite the Lagrangian in the canonical form. In an effective theory, we can perform non-linear field redefinitions due to the presence of higher dimensional operators. This invariance gives rise to a rule to remove redundant terms from the effective Lagrangian and leads to the construction of ``on-shell" effective theory~\cite{Georgi:1991ch}. 

One way of removing the redundancies with higher derivatives of operators involving the Higgs field $H$ is to redefine the field in a perturbative manner~\cite{Georgi:1991ch,Corbett:2021eux,Barzinji:2018xvu,Criado:2018sdb}. For example, to remove the term $\,\boldsymbol{\xi^{(6)}_{1}(H^{\dagger}H)(\mathcal{D}^2 H^{\dagger}H+H^{\dagger}\mathcal{D}^2 H)}$ in Eq.~\eqref{eq:Lagrangian}, we can use the redefinition $H\to H+\xi^{(6)}_{1}(H^{\dagger}H)H$, in which case the redundancy at the level of $\mathcal{O}(\xi^{(6)}_{1})$ will be removed. \sr{Subsequently, it will give rise to higher dimensional operator at $\mathcal{O}\big((\xi^{(6)}_{1})^2\big)$.}

Now the same outcome can be achieved by employing the EOM judiciously.  We compute the classical EOM for the Higgs considering all possible structures up to dimension-six from Eq.~\eqref{eq:Lagrangian}
\begin{multline}
\label{eq:EOM_dim6}
		\mathcal{D}^2 H = 
		\underline{-\,2\lambda'(H^{\dagger}H)H-\mathcal{Y}}
		+3\,\zeta^{(6)}_{1}(H^{\dagger}H)^2H
		+\zeta^{(6)}_3(\mathcal{D}_{\mu}H^{\dagger}H)\mathcal{D}^{\mu}H\\
		+\xi^{(6)}_1 (\mathcal{D}^2 H^{\dagger} H+H^{\dagger}\mathcal{D}^2 H)H
		+\xi^{(6)}_1(H^{\dagger}H)\,\mathcal{D}^2 H 
		-\zeta^{(6)}_3 \,\mathcal{D}_{\mu} \big[(H^{\dagger}\mathcal{D}_{\mu}H)H\big]\\
		+2\zeta_{2}^{(6)} H\square(H^{\dagger}H)+\xi_1^{(6)}\,\mathcal{D}^2[(H^{\dagger}H)H]
		+\zeta_{4}^{(6)} \,H(B_{\mu\nu}B^{\mu\nu})\\
		+\zeta_5^{(6)} H(W_{\mu\nu}^IW^{I,\mu\nu})
		+\,\zeta_{6}^{(6)}(\tau^I H)(B_{\mu\nu}W^{I,\mu\nu})+\zeta^{(6)}_7H(G^{a}_{\mu\nu}G^{a\mu\nu})\\
		+\boxed{i\,\zeta_{8,1}^{(6)}\mathcal{D}_{\mu}\big[H\,(\overline{\psi}\gamma^{\mu}\psi)\big]}
		+i\,\zeta_{8,1}^{(6)} (\mathcal{D}_{\mu}H)(\overline{\psi}\gamma^{\mu}\psi)
		+\boxed{i\,\zeta_{8,2}^{(6)}\mathcal{D}_{\mu}^{I}\big[H\,(\overline{\psi}\gamma^{\mu}\tau^{I}\psi)\big]}\\
		+i\,\zeta_{8,2}^{(6)} \,(\mathcal{D}_{\mu}^{I}H)(\overline{\psi}\gamma^{\mu}\tau^{I}\psi)+\boxed{\xi^{(6)}_2\,\mathcal{D}^{2}(\overline{\psi}\,\psi)}\,.
\end{multline}
\noindent Here, $\lambda^{'} = \lambda - \widetilde{\lambda}$ with $\lambda$ being the SM Higgs quartic coupling in the renormalisable Lagrangian. $\widetilde{\lambda}$ is the direct contribution to the former, obtained from integrating out the heavy field as shown in Eq.~\eqref{eq:Lagrangian}. The underlined part on the right-hand side of the Eq.~\eqref{eq:EOM_dim6} denotes the contribution from the renormalisable part of the Lagrangian $ (\mathcal{L}_{\text{SM}}^{(4)}) $, which is considered as the first-order term in the EOM. The remainder arises from the effective operators at dimension-six and is considered second-order terms~\cite{Manohar:2018aog}. We can think of Eq.~\eqref{eq:EOM_dim6} as some special field redefinition and directly employ the first-order terms to remove the redundancies in Eq.~\eqref{eq:Lagrangian}, but this will not generate any higher dimensional structures. This is the common practice to obtain the complete basis at dimension six. Working with the second-order terms is a non-trivial task, substituting it directly into Eq.~\eqref{eq:Lagrangian} to obtain a contribution to higher dimension operators could lead to incompleteness as pointed out in~\cite{Criado:2018sdb}: The missing contributions can be encapsulated by including a term, $ (1/2)\big(\xi^{(6)}_{1}(H^{\dagger}H)\big)^2 \delta^2 \mathcal{L}/\delta H^\dagger \delta H  $. Calculating this term from Eq.~\eqref{eq:Lagrangian} we obtain the following contribution:

\begin{multline}
\label{eq:extra-term}
		\mathcal{L}' = \big(\xi_{1}^{(6)}\big)^2 \,\mathcal{D}_{\mu}\big[(H^{\dagger}H)H^{\dagger}\big]\,\mathcal{D}^{\mu}\big[(H^{\dagger}H)H\big]\,-6 \,\big(\xi_{1}^{(6)}\big)^2 \lambda' (H^{\dagger}H)^4 \\
		= -2\,\big(\xi_{1}^{(6)}\big)^2\lambda'(H^{\dagger}H)^4 + \big(\xi_{1}^{(6)}\big)^2 (H^{\dagger}H)^2\,(H^{\dagger}\mathcal{Y}+\mathcal{Y}^{\dagger}H)-\big(\xi_{1}^{(6)}\big)^2 (H^{\dagger}H)^2(\mathcal{D}_{\mu}H^{\dagger}\mathcal{D}^{\mu}H).
\end{multline}
Since our primary concern is the redundant operators involving the Higgs field, the boxed structures in the Eq.~\eqref{eq:EOM_dim6} containing the derivative of fermion fields can be reduced to other structures by applying the first-order fermionic EOM.

We are now ready to implement the methodology discussed above to compute the dimension-eight coefficients from dimension-six operators.

\subsection{Impact of dimension-six structures on dimension-eight coefficients}
\label{subsec:dim-6-tree-Lag}
It is a common practice to employ the first-order EOM, i.e., the classical equation of motion obtained from the renormalisable Lagrangian, to transform the operators from one basis to another at a given mass dimension. Here we extend this strategy to generate higher-order terms in the effective Lagrangian. The contribution to the WCs arising from the EOM substitution considering second-order terms will be important. In Table~\ref{Tab:effective_ops_to_dim6} we provide the contribution to dimension-eight operators coming from the dimension-six Lagrangian. The operator structures are shown in appendix~\ref{app:op-structures}.
 
	\begin{table}[!t]
	\centering \scriptsize
	\renewcommand{\arraystretch}{2.2}
     \resizebox{\textwidth}{!}{\begin{tabular}{|c|c|c|c|}
		\hline
		\textsf{\quad Operator}$\quad$&
		\textsf{\quad Wilson coefficients}$\quad$&
		\textsf{\quad Operator}$\quad$&
		\textsf{\quad Wilson coefficients}$\quad$\\
		\hline

		$\mathcal{O}_{H^6\mathcal{D}^2,1}^{(8)}$&
		$\big(\xi_{1}^{(6)}\big)^2+8\zeta_{2}^{(6)}\xi_{1}^{(6)}-\zeta_{3}^{(6)}\xi_{1}^{(6)}$&
		$\mathcal{O}_{H^6\mathcal{D}^2,2}^{(8)}$&
		$4\zeta_3^{(6)}\xi_1^{(6)}$\\
		\hline
		
		$\mathcal{O}_{\psi^2H^4\mathcal{D},1}^{(8)}$&
		$i\,\zeta_{8,1}^{(6)}\,\xi_{1}^{(6)}$&
		$\mathcal{O}_{\psi^2H^4\mathcal{D},2}^{(8)}$&
		$i\,\zeta_{8,2}^{(6)}\,\xi_1^{(6)}$\\
		\hline
		
		$\mathcal{O}_{\psi^4\mathcal{D}H,1}^{(8)}$&
		$i\,\zeta_{8,1}^{(6)}\,\xi_{2}^{(6)}$&
		$\mathcal{O}_{\psi^4\mathcal{D}H,2}^{(8)}$&
		$i\,\zeta_{8,2}^{(6)}\,\xi_2^{(6)}$\\
		\hline
		
		$\mathcal{O}_{H^4B^2}^{(8)}$&
		$2\zeta_{4}^{(6)}\,\xi_{1}^{(6)}$&
		$\mathcal{O}_{H^4W^2}^{(8)}$&
		$2\zeta_{5}^{(6)}\xi_{1}^{(6)}$\\
		\hline
		
		$\mathcal{O}_{H^4WB}^{(8)}$&
		$2\zeta_{6}^{(6)}\xi_1^{(6)}$&
		$\mathcal{O}_{H^4G^2}^{(8)}$&
		$2\zeta_{7}^{(6)}\xi_1^{(6)}$\\
		\hline
		
		$\mathcal{O}_{\psi^2B^2H}^{(8)}$&
		$\zeta_{4}^{(6)}\xi_2^{(6)}$&
		$\mathcal{O}_{\psi^2W^2H}^{(8)}$&
		$\zeta_{5}^{(6)}\xi_2^{(6)}$\\
		\hline
		
		$\mathcal{O}_{\psi^2WBH}^{(8)}$&
		$\zeta_6^{(6)}\xi_2^{(6)}$&
		$\mathcal{O}_{\psi^2G^2H}^{(8)}$&
		$\zeta_7^{(6)}\xi_2^{(6)}$
		\\
		\hline	
		
		\multirow{3}{*}{$\mathcal{O}_{\psi^2H^5}^{(8)}$}&
		$\big(-4\big(\xi_1^{(6)}\big)^2-4\zeta_2^{(6)}\xi_1^{(6)}+\frac{1}{2}\zeta_3^{(6)}\xi_1^{(6)}\big)\,Y_{\text{SM}}$&
		\multirow{2}{*}{$\mathcal{O}_{H}^{(8)}$}&
		$6\zeta_1^{(6)}\xi_1^{(6)}-6\big(\xi_1^{(6)}\big)^2\lambda'-4\lambda'(4(\xi_1^{(6)})^2$\\
		
		&
		$+3\zeta_{1}^{(6)}\xi_{2}^{(6)}-12\lambda'\xi_1^{(6)}\xi_2^{(6)}+2\lambda'\xi_{2}^{(6)}\zeta_{3}^{(6)}$&
		&$+4\zeta_2^{(6)}\xi_{1}^{(6)}-\frac{1}{2}\zeta_3^{(6)}\xi_{1}^{(6)})$\\

		&
		$-8\lambda'\xi_{2}^{(6)}\zeta_{2}^{(6)}$&
		&\\
		\hline
		
		$\mathcal{O}_{\psi^2H^3\mathcal{D}^2,1}^{(8)}$&
		$+4\xi_2^{(6)}\zeta_{2}^{(6)}+2\xi_1^{(6)}\xi_{2}^{(6)}$&
		$\mathcal{O}_{\psi^2H^3\mathcal{D}^2,2}^{(8)}$&
		$-\xi_{2}^{(6)}\zeta_{3}^{(6)}$\\
		\hline

		$\mathcal{O}_{\psi^4H^2,1}^{(8)}$&
		$(-4\xi_1^{(6)}\xi_2^{(6)}-4\zeta_2^{(6)}\xi_2^{(6)}+2\zeta_3^{(6)}\xi_2^{(6)})\,Y_{\text{SM}}$&
		$\mathcal{O}_{\psi^4H^2,2}^{(8)}$&
	  $(-3\xi_1^{(6)}\xi_2^{(6)}-2\xi_2^{(6)}\zeta_{2}^{(6)})\,Y_{\text{SM}}$\\
		\hline		
	\end{tabular}}
	\caption{\small Contributions to the dimension-eight operators from the dimension-six structures after implementing the EOM. Here $Y_{\text{SM}}$ denotes the SM Yukawa coupling. We have suppressed the flavour indices without any loss of generality.}
	\label{Tab:effective_ops_to_dim6}
\end{table}

As the process of integrating out heavy fields becomes more complicated at higher operator dimensions, our method of generating WCs from lower dimension ones becomes economical. Following the expressions shown in Table~\ref{Tab:effective_ops_to_dim6}, one can quickly work out the dimension-eight contribution without explicitly performing the matching at that order. In the following subsection, we compute the complete basis at dimension eight.

\subsection{Removing redundancies at dimension-eight}
\label{subsec:dim-8-tree-Lag}
We consider all (redundant and non-redundant) structures that only involve $H$ and its derivatives at dimension-eight that can arise directly after integrating out at tree-level.
\begin{multline}
\label{eq:Lagrangian_dim8}
\mathcal{L}_{\text{eff}}^{(8)} = 
	\zeta^{(8)}_1 (H^{\dagger}H)^4+\zeta^{(8)}_2 (H^{\dagger}H)(H^{\dagger}\mathcal{D}_{\mu}H\,\mathcal{D}^{\mu}H^{\dagger}H)
	+\zeta^{(8)}_3(H^{\dagger}H)^2(\mathcal{D}_{\mu}H^{\dagger}\mathcal{D}^{\mu}H)\\
	+\zeta^{(8)}_4\,(\mathcal{D}_{\mu}H^{\dagger}\mathcal{D}^{\nu}H)(\mathcal{D}^{\nu}H^{\dagger}\mathcal{D}_{\mu}H)
	+\zeta^{(8)}_5\,(\mathcal{D}_{\mu}H^{\dagger}\mathcal{D}^{\nu}H)(\mathcal{D}^{\mu}H^{\dagger}\mathcal{D}_{\nu}H)\\
	+\zeta^{(8)}_6\,(\mathcal{D}_{\mu}H^{\dagger}\mathcal{D}^{\mu}H)(\mathcal{D}_{\nu}H^{\dagger}\mathcal{D}^{\nu}H)
	+\boldsymbol{\
 }\\
	+\boldsymbol{\xi^{(8)}_2(\mathcal{D}_{\mu}H^{\dagger}\mathcal{D}^{\mu}H)(\mathcal{D}^2 H^{\dagger} H+H^{\dagger}\mathcal{D}^2 H)}	
	+\boldsymbol{\xi^{(8)}_3\big[(\,\mathcal{D}_{\mu}H^{\dagger}H)(\mathcal{D}^2H^{\dagger}\mathcal{D}^{\mu}H)+\textbf{h.c.}\big]}\\
	+\boldsymbol{\xi^{(8)}_4\,\big[(\mathcal{D}^2H^{\dagger}H)(\mathcal{D}^2H^{\dagger}H)+\textbf{h.c.}\big]}
	+\boldsymbol{\xi^{(8)}_5\,\big[(\mathcal{D}_{\mu}H^{\dagger}H)(\mathcal{D}_{\mu}H^{\dagger}\mathcal{D}^2H)+\textbf{h.c.}\big]}\\
	+\boldsymbol{\xi^{(8)}_6\,(\mathcal{D}^2H^{\dagger}\mathcal{D}^2H)(H^{\dagger}H)}
	+\boldsymbol{\xi^{(8)}_7\,(H^{\dagger}\mathcal{D}^2H)(\mathcal{D}^2H^{\dagger}H)}.	
\end{multline}
The redundant structures, written in bold in Eq.~\eqref{eq:Lagrangian_dim8}, can be expressed in terms of the non-redundant basis structures in the following manner \footnote{These operators can be related to the structures shown in Ref.~\cite{Chala:2021cgt}.}:
\begin{alignat}{3}
\label{eq:redundant_to_nonredundant} 
	\xi_1^{(8)}(H^{\dagger}H)^2(\mathcal{D}^2H^{\dagger}H+H^{\dagger}\mathcal{D}^2H)=&-4\lambda'\xi_1^{(8)}(H^{\dagger}H)^4-\xi_1^{(8)}\big[(H^{\dagger}H)^2(\mathcal{Y}^{\dagger}H)+\text{h.c.}\big]\nonumber\\
	\xi_2^{(8)}(\mathcal{D}_{\mu}H^{\dagger}\mathcal{D}^{\mu}H)(\mathcal{D}^2 H^{\dagger} H+H^{\dagger}\mathcal{D}^2 H)=&-4\lambda'\xi_{2}^{(8)}(H^{\dagger}H)^2(\mathcal{D}_{\mu}H^{\dagger}\mathcal{D}^{\mu}H)\nonumber\\ &-\xi_{2}^{(8)}\big[(H^{\dagger}\mathcal{Y})(\mathcal{D}_{\mu}H^{\dagger}\mathcal{D}^{\mu}H)+\text{h.c.}\big],\nonumber\\
	\xi_{3}^{(8)}\big[(\,\mathcal{D}_{\mu}H^{\dagger}H)(\mathcal{D}^2H^{\dagger}\mathcal{D}^{\mu}H)+\text{h.c.}\big]=&-4\lambda'\xi_{3}^{(8)}(H^{\dagger}H)(H^{\dagger}\mathcal{D}_{\mu}H)(\mathcal{D}^{\mu}H^{\dagger}H)\nonumber\\&-\xi_{3}^{(8)}\big[(H^{\dagger}\mathcal{Y})(\mathcal{D}_{\mu}H^{\dagger}\mathcal{D}^{\mu}H)+\text{h.c.}\big],\nonumber\\
	\xi_4^{(8)}\,\big[(\mathcal{D}^2H^{\dagger}H)(\mathcal{D}^2H^{\dagger}H)+\text{h.c.}\big]=&\,8\lambda'^{2}\xi_4^{(8)}(H^{\dagger}H)^4+4\lambda'\xi_4^{(8)}\big[(H^{\dagger}H)^2(\mathcal{Y}^{\dagger}H)+\text{h.c.}\big]\nonumber\\&+\xi_4^{(8)}\big[(\mathcal{Y}^{\dagger}H)(\mathcal{Y}^{\dagger}H)+\text{h.c.}\big],\nonumber\\
	\xi_5^{(8)}\big[(\mathcal{D}_{\mu}H^{\dagger}H)(\mathcal{D}_{\mu}H^{\dagger}\mathcal{D}^2H)+\text{h.c.}\big]=&-4\lambda'\xi_5^{(8)}(H^{\dagger}H)(H^{\dagger}\mathcal{D}_{\mu}H)(\mathcal{D}^{\mu}H^{\dagger}H)\nonumber\\&-\xi_5^{(8)}\big[(\mathcal{D}_{\mu}H^{\dagger}H)(\mathcal{D}^{\mu}H^{\dagger}\mathcal{Y})+\text{h.c.}\big],\nonumber\\
	\xi_6^{(8)}\,(\mathcal{D}^2H^{\dagger}\mathcal{D}^2H)(H^{\dagger}H)=&4\lambda'^{2}\xi_6^{(8)}(H^{\dagger}H)^4+2\lambda'\xi_6^{(8)}\big[(H^{\dagger}H)^2(\mathcal{Y}^{\dagger}H)+\text{h.c.}\big]\nonumber \\ &+\xi_6^{(8)}(H^{\dagger}H)(\mathcal{Y}^{\dagger}\mathcal{Y}),\nonumber\\
	\xi_7^{(8)}\,(H^{\dagger}\mathcal{D}^2H)(\mathcal{D}^2H^{\dagger}H)=&4\lambda'^{2}\xi_7^{(8)}(H^{\dagger}H)^4+2\lambda'\xi_7^{(8)}\big[(H^{\dagger}H)^2(\mathcal{Y}^{\dagger}H)+\text{h.c.}\big]\nonumber\\&+\xi_7^{(8)}(H^{\dagger}H)(\mathcal{Y}^{\dagger}\mathcal{Y}).
\end{alignat}		
In Table~\ref{Tab:effective_ops_to_dim8}, we present the coefficients in the non-redundant basis. 

\begin{figure}
    \centering
    \includegraphics[scale=0.6]{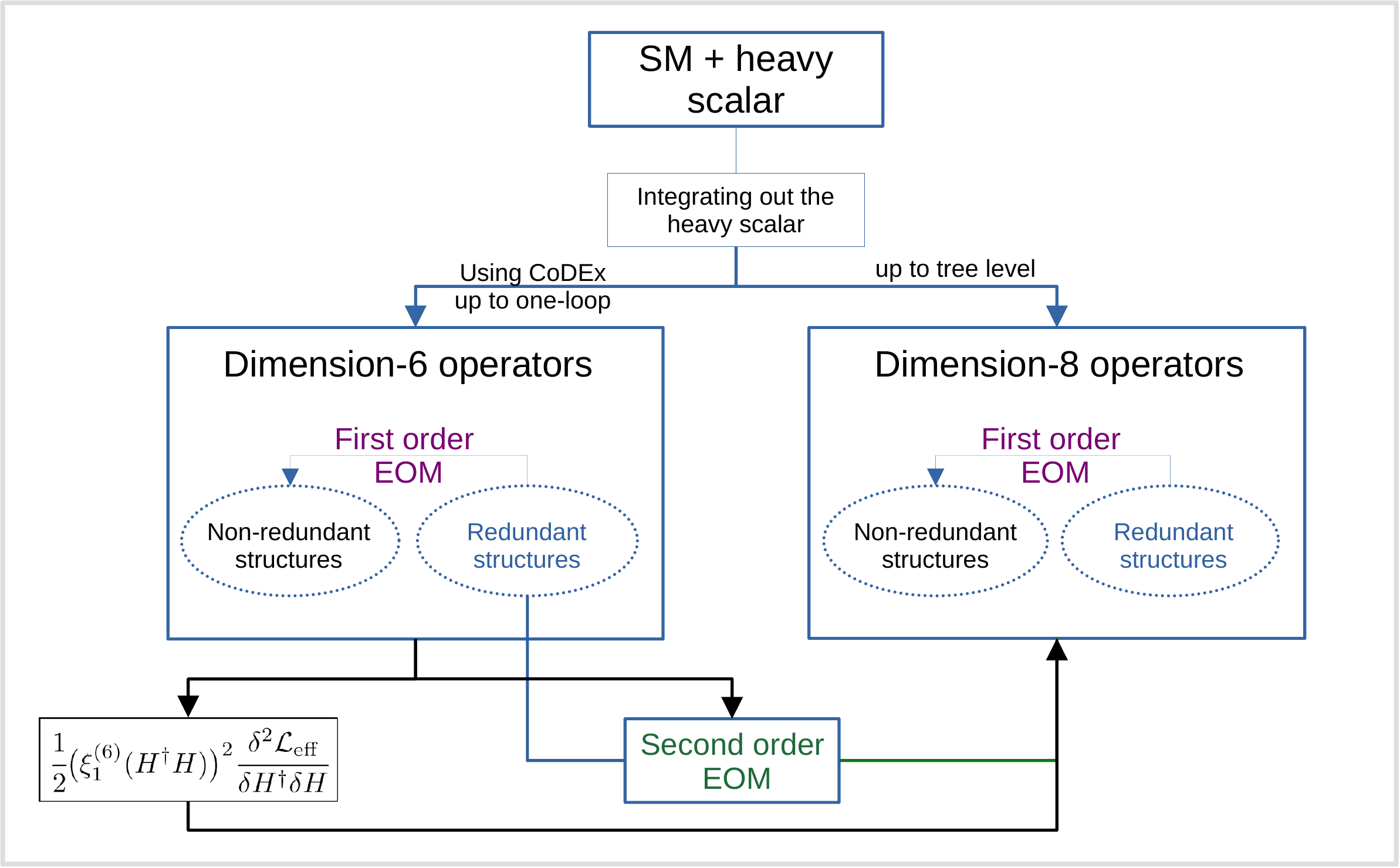}
    \caption{Flow chart depicting the algorithmic approach considered to compute matching coefficients for both dimension-six and dimension-eight operators. Here, ``First order EOM" and ``Second order EOM" are formulated from the renormalizable and the dimension-six parts of the SM Lagrangian respectively.}
    \label{fig:flowchart}
\end{figure}

\begin{table}[!h]
			\centering \scriptsize
			\renewcommand{\arraystretch}{2.6}
			\resizebox{\textwidth}{!}{\begin{tabular}{|c|c|c|c|}
				\hline
				\textsf{\quad Operator}$\quad$&
				\textsf{\quad Wilson coefficients}$\quad$&
				\textsf{\quad Operator}$\quad$&
				\textsf{\quad Wilson coefficients}$\quad$\\
				\hline

				\multirow{2}{*}{$\mathcal{O}_{H}^{(8)}$}&
				$\zeta_1^{(8)}-4\lambda'\xi_1^{(8)}+8\lambda'^{2}\xi_{4}^{(8)}$&
				\multirow{2}{*}{$\mathcal{O}_{\psi^2H^5}^{(8)}$}&
				$(-\xi_1^{(8)}+4\lambda'\xi_{4}^{(8)}+2\lambda'\xi_{6}^{(8)}$\\

				&
				$+4\lambda'^{2}\xi_{6}^{(8)}+4\lambda'^{2}\xi_{7}^{(8)}$&
				&$+2\lambda'\xi_{7}^{(8)})\,Y_{\text{SM}}$\\
				\hline

				$\mathcal{O}_{H^6\mathcal{D}^2,1}^{(8)}$&
				$\zeta_3^{(8)}-4\lambda'\xi_{2}^{(8)}$&
				$\mathcal{O}_{H^6\mathcal{D}^2,2}^{(8)}$&
				$\zeta_{2}^{(8)}-4\lambda'\xi_{3}^{(8)}-4\lambda'\xi_5^{(8)}$\\
				\hline

				$\mathcal{O}_{\psi^2H^3
					\mathcal{D}^2,1}^{(8)}$&
				$(-\xi_{2}^{(8)}-\xi_{3}^{(8)})\,Y_{\text{SM}}$&
				$\mathcal{O}_{\psi^2H^3
					\mathcal{D}^2,2}^{(8)}$&
				$(-\xi_{5}^{(8)})\,Y_{\text{SM}}$\\
				\hline

				$\mathcal{O}_{\psi^4H^2,1}^{(8)}$&
				$(\xi_{6}^{(8)}+\xi_{7}^{(8)})\,Y_{\text{SM}}^2$&
				$\mathcal{O}_{\psi^4H^2,2}^{(8)}$&
				$(\xi_{4}^{(8)})\,Y_{\text{SM}}^2$\\
				\hline
				
			\end{tabular}}
			\caption{\small Matching contributions to the non-redundant dimension-eight operators from the dimension-eight structures after implementing the EOM.
			Here $Y_{\text{SM}}$ denotes the SM Yukawa coupling. We have suppressed the flavour indices without any loss of generality and continued throughout the rest of the paper.}
			\label{Tab:effective_ops_to_dim8}
		\end{table} 
Before cross-checking the proposed method in the next section, we summarise our framework in the flowchart depicted in Fig.~\ref{fig:flowchart}. This work considers SM extensions of only a single heavy scalar. \texttt{CoDEx}~\cite{Bakshi:2018ics} has been used to generate the operators and the WCs in the \emph{SILH set} up to one-loop at dimension-six. We compute the EOM (only for the Higgs field), including contributions from dimension-six operators; we substitute the EOM in the redundant structures. The first-order terms transform the redundant structure of dimension-six to non-redundant structures, while the second-order terms generate dimension-eight operators. To compensate for the missing contribution that renders the EOM equivalent to a field redefinition, we need to add a term proportional to the second-order derivative of the effective action. Furthermore, we calculate dimension-eight operators by integrating out the heavy field at the tree level, which gives rise to the leading effects at dimension eight. We then substitute the first-order terms in the EOM to convert them into a complete basis and combine all these contributions to obtain the complete matching result. 

The following section applies this to reproduce the known results for the real triplet extension of the SM to validate our methodology.

\section{Cross-validation of the method}
\label{sec:x-validation}
To cross check our approach, we first turn to the
example case of a real triplet scalar ($\Phi$) SM extension. The BSM part of the Lagrangian reads
\begin{equation}
\label{eq:real_triplet_Lagrangian} 
			\mathcal{L}_{\Phi} = \frac{1}{2}(\mathcal{D}_{\mu}\Phi^a)(\mathcal{D}^{\mu}\Phi^a)-\frac{1}{2}m_{\Phi}^2\Phi^a \Phi^a+2kH^{\dagger}\tau^a H \Phi^a-\eta (H^{\dagger}H)\Phi^a \Phi^a -\frac{1}{4}\lambda_{\Phi}(\Phi^a \Phi^a)^2.
\end{equation}
Integrating out the heavy scalar leads to some correction to the renormalisable term $(H^{\dagger}H)^2$ as discussed in Eq.~\eqref{eq:Lagrangian}, the coefficient $\widetilde{\lambda}$ for this case is: $\widetilde{\lambda}=k^2/(2m_{\Phi}^2)$.
    
The SILH dimension-six coefficients have been tabulated in Table~\ref{Tab:real_triplet_SILH_coefficients}. The one-loop contribution to the matching can be categorised into two different classes: the contribution arising from integrating out scalars from purely heavy loops has been highlighted in blue, and terms from heavy-light mixed loops are shown in red.   

\begin{table}[!t]
	\centering
	\renewcommand{\arraystretch}{2.6} \scriptsize
	\resizebox{\textwidth}{!}{\begin{tabular}{|c|c|c|c|}
		\hline
		\textsf{\quad SILH Op.}$\quad$&
		\textsf{\quad Wilson Coefficients}$\quad$&
		\textsf{\quad SILH Op.}$\quad$&
		\textsf{\quad Wilson Coefficients}$\quad$\\
		\hline
		
	    \multirow{3}{*}{$\mathcal{O}_{6}$}&
	    $-\frac{\eta k^2}{m_{\Phi}^4}-\biggl \{\frac{\eta^3}{8m_{\Phi}^2\pi^2}-\frac{5\eta k^2\lambda_{\Phi}}{8m_{\Phi}^4\pi^2}\biggr \}+\biggl [\frac{13\eta^2k^2}{8m_{\Phi}^4\pi^2}$
		&\multirow{2}{*}{$\mathcal{O}_{H}$}
		&$\biggl \{\frac{\eta^2}{16m_{\Phi}^2\pi^2} \biggr \}-\biggl [\frac{3\eta k^2}{8m_{\Phi}^4\pi^2}-\frac{9k^4}{32m_{\Phi}^6\pi^2}$
        \\
        
        &
        $+\frac{47\eta k^4}{16m_{\Phi}^6\pi^2}+\frac{19k^6}{16m_{\Phi}^8\pi^2}-\frac{2\eta k^2\lambda}{m_{\Phi}^4\pi^2}-\frac{2k^4\lambda}{m_{\Phi}^6\pi^2}$&
        &
        $+\frac{5k^2\lambda}{16m_{\Phi}^4\pi^2} \biggr ]$
        \\
        
        &
        $+\frac{11k^2\lambda^2}{16m_{\Phi}^4\pi^2}-\frac{5k^4\lambda_{\Phi}}{16m_{\Phi}^6\pi^2} \biggr ]$&
        &
        \\
		\hline

		\multirow{2}{*}{$\mathcal{O}_R$}&
		$\frac{2k^2}{m_{\Phi}^4}+\biggl \{\frac{5k^2\lambda_{\Phi}}{4m_{\Phi}^4\pi^2}\biggr \}-\biggl [\frac{21\eta k^2}{16m_{\Phi}^4\pi^2}$&
		\multirow{2}{*}{$\mathcal{O}_{T}$}&
		$\frac{k^2}{m_{\Phi}^4}+\biggl \{\frac{5k^2\lambda_{\Phi}}{8m_{\Phi}^4\pi^2} \biggr \}-\biggl [\frac{\eta k^2}{2m_{\Phi}^4\pi^2}$
		\\

		&
		$-\frac{21 k^4}{32m_{\Phi}^6\pi^2}+\frac{25k^2\lambda}{32m_{\Phi}^4\pi}\biggr ]$&
		&
		$+\frac{k^4}{32m_{\Phi}^6\pi^2}+\frac{3k^2\lambda}{32m_{\Phi}^4\pi^2}\biggr]$
		\\
		\hline

		$\mathcal{O}_{WW}$&
		$\biggl\{\frac{\eta}{96m_{\Phi}^2\pi^2}\biggr\}+\biggl[\frac{25k^2}{768m_{\Phi}^4\pi^2}\biggr]$&
		$\mathcal{O}_{2W}$&
		$\biggl \{\frac{g_{_W}^2}{480m_{\Phi}^2\pi^2} \biggr\}$
		\\
		\hline

        $ \mathcal{O}_{WB}$&
        $\biggl [-\frac{k^2}{128m_{\Phi}^4\pi^2} \biggr ]$&
        $\mathcal{O}_{BB}$&
        $\biggl [\frac{3k^2}{256m_{\Phi}^4\pi^2} \biggr]$
        \\
        \hline

        $\mathcal{O}_{W}$&
        $\biggl [-\frac{k^2}{288m_{\Phi}^4\pi^2} \biggr]$&
        $\mathcal{O}_{B}$&
        $\biggl [-\frac{7k^2}{96m_{\Phi}^4\pi^2} \biggr]$
        \\
        \hline
        
	\end{tabular}}
	\caption{\small Dimension-six SILH Wilson coefficients relevant for integrating out the \textbf{real-triplet scalar} of Eq.~\eqref{eq:real_triplet_Lagrangian} at one-loop. The terms within braces $(\{\})$ denote the contribution from pure heavy loops, whereas the brackets $(\left[ \,\right])$ mark the contribution from light-heavy mixed loops.}
	\label{Tab:real_triplet_SILH_coefficients}
\end{table}

We compute the Green's set-like coefficients first following the relations provided in Table~\ref{Tab:SILH_to_mixed} and derive the corrections to dimension-eight from dimension-six structures as shown in Table~\ref{Tab:effective_ops_to_dim6}. These contributions are computed considering only the tree-level part of the SILH coefficients. Thus, they are on the same footing as the direct dimension-eight contributions; they are tabulated separately in Table~\ref{Tab:real_triplet_dim6_to_dim8}. 
%
	\begin{table}[!t]
	\centering
	\renewcommand{\arraystretch}{2.6} \scriptsize
	\begin{tabular}{|c|c|c|c|}
		\hline
		\textsf{\quad Operator}$\quad$&
		\textsf{\quad Wilson coefficients}$\quad$&
		\textsf{\quad Operator}$\quad$&
		\textsf{\quad Wilson coefficients}$\quad$\\
		\hline

		$\mathcal{O}_{H}^{(8)}$&
		$\frac{5k^6}{m_{\Phi}^{10}}+\frac{6k^4\eta}{m_{\Phi}^8}-\frac{10k^4\lambda}{m_{\Phi}^8}$
		&$\mathcal{O}_{\psi^2H^5}^{(8)}$&
		$-\frac{k^4}{m_{\Phi}^8}\,Y_{\text{SM}}$
		\\
		\hline
		
		$\mathcal{O}_{H^6\mathcal{D}^{2},1}^{(8)}$&
		$-\frac{5 k^4}{m_{\Phi}^8}$
       &$\mathcal{O}_{H^6\mathcal{D}^2,2}^{(8)}$&
        $ \frac{8 k^4}{m_{\Phi}^8}$
        \\
        \hline

	\end{tabular}
	\caption{\small Contributions to dimension-eight operators from dimension-six structures when integrating out the heavy \textbf{real triplet scalar} of Eq.~\eqref{eq:real_triplet_Lagrangian}. Here only the tree-level matching of the dimension-six structures have been considered while computing the results.}
	\label{Tab:real_triplet_dim6_to_dim8}
\end{table}
%
The direct contributions at dimension-eight after  ``integrating out" can be captured by a total of seven coefficients as specified in Eq.~\eqref{eq:Lagrangian_dim8}. The values for the coefficients are
      \begin{eqnarray}\label{eq:real-triplet-dim8-coefficients}
          \zeta_1^{(8)}&=&\big(\frac{2\eta^2k^2}{m_{\Phi}^6}-\frac{k^4}{4m_{\Phi}^8}\lambda_{\Phi}\big),\hspace{3mm} \zeta_2^{(8)}=\frac{8\eta k^2}{m_{\Phi}^6},\hspace{3mm}\zeta_3^{(8)}=-\frac{4\eta k^2}{m_{\Phi}^6},\nonumber\\
          \zeta_4^{(8)}&=&\frac{4k^2}{m_{\Phi}^6},\hspace{3mm}\zeta_5^{(8)}=0,\hspace{3mm}\zeta_6^{(8)}=-\frac{2k^2}{m_{\Phi}^6},\hspace{3mm}\xi_1^{(8)}=\frac{2\eta k^2}{m_{\Phi}^6},\nonumber\\
          \xi_2^{(8)}&=&-\frac{2k^2}{m_{\Phi}^6}, \hspace{4mm}\xi_3^{(8)}=\frac{4k^2}{m_{\Phi}^6},\hspace{4mm}\xi_4^{(8)}=\frac{k^2}{2m_{\Phi}^6},\hspace{4mm}\xi_5^{(8)}=0,\nonumber\\
          \xi_6^{(8)}&=&\frac{2k^2}{m_{\Phi}^6},\hspace{4mm}\xi_7^{(8)}=-\frac{k^2}{m_{\Phi}^6}\,.
      \end{eqnarray}
In Table~\ref{Tab:real_triplet_dim8_to_dim8}, we show the direct contributions to dimension-eight structures removing redundancies at dimension-eight following the relations given in Table~\ref{Tab:effective_ops_to_dim8}.
%
\begin{table}[!t]
	\centering
	\renewcommand{\arraystretch}{2.6} \scriptsize
	\begin{tabular}{|c|c|c|c|}
		\hline
		\textsf{\quad Operator}$\quad$&
		\textsf{\quad Wilson coefficients}$\quad$&
		\textsf{\quad Operator}$\quad$&
		\textsf{\quad Wilson coefficients}$\quad$\\
		\hline
		
	\multirow{2}{*}{$\mathcal{O}_{H}^{(8)}$}&
		$\frac{2 k^6}{m_{\Phi}^{10}}+\frac{4 \eta  k^4}{m_{\Phi}^8}-\frac{8 \lambda 
   k^4}{m_{\Phi}^8}-\frac{k^4 \lambda_{\Phi} }{4 m_{\Phi}^8}$&
		$\mathcal{O}_{\psi^2H^5}^{(8)}$&
		$\big(-\frac{2 k^4}{m_{\Phi}^8}-\frac{2 \eta  k^2}{m_{\Phi}^6}+\frac{4 \lambda  k^2}{m_{\Phi}^6}\big)Y_{\text{SM}}$
		\\
		
		&
		$+\frac{2 \eta ^2 k^2}{m_{\Phi}^6}-\frac{8
   \eta  \lambda  k^2}{m_{\Phi}^6}+\frac{8 \lambda ^2 k^2}{m_{\Phi}^6}$
		&
		&
		\\
		\hline

		$\mathcal{O}_{H^6\mathcal{D}^2,1}^{(8)}$&
		$-\frac{4 k^4}{m_{\Phi}^8}-\frac{4 \eta  k^2}{m_{\Phi}^6}+\frac{8 \lambda  k^2}{m_{\Phi}^6}$
		&$\mathcal{O}_{H^6\mathcal{D}^2,2}^{(8)}$
		&
		$\frac{8 k^4}{m_{\Phi}^8}+\frac{8 \eta  k^2}{m_{\Phi}^6}-\frac{16 \lambda  k^2}{m_{\Phi}^6}$
		\\
		\hline

	\end{tabular}
	\caption{\small Contributions to dimension-eight operators from dimension-eight structures when integrating out the heavy \textbf{real triplet scalar} of Eq.~\eqref{eq:real_triplet_Lagrangian}.}
	\label{Tab:real_triplet_dim8_to_dim8}
\end{table}
\begin{table}[!t]
	\centering
	\renewcommand{\arraystretch}{2.6} \scriptsize
	\begin{tabular}{|c|c|c|c|}
		\hline
		\textsf{\quad Operator}$\quad$&
		\textsf{\quad Wilson coefficients}$\quad$&
		\textsf{\quad Operator}$\quad$&
		\textsf{\quad Wilson coefficients}$\quad$\\
		\hline

		$\mathcal{O}_{H}^{(8)}$&
		$\frac{7k^6}{m_{\Phi}^{10}}+\frac{2k^2(2\lambda-\eta)^2}{m_{\Phi}^6}+\frac{(40\eta-72\lambda-\lambda_{\Phi})k^4}{4m_{\Phi}^8}$&$\mathcal{O}_{H^6\mathcal{D}^2,1}^{(8)}$
		&$-\frac{k^4}{m_{\Phi}^8}$
		\\

		\hline

		$\widetilde{\mathcal{O}}_{H^6\mathcal{D}^2,2}^{(8)}$&
		$\frac{(4\eta-8\lambda)k^2}{m_{\Phi}^6}+\frac{8k^4}{m_{\Phi}^8}$
		&$\mathcal{O}_{\psi^2H^5}^{(8)}$&
		$Y_{\text{SM}}\big(\frac{-3k^4-2m^2k^2(\eta-2\lambda)}{m_{\Phi}^8}\big)$\\

		\hline

	\end{tabular}
	\caption{\small Total tree-level matching of the dimension-eight coefficients for the \textbf{real-triplet scalar} extension of SM of Eq.~\eqref{eq:real_triplet_Lagrangian}.}
	\label{Tab:real_triplet_dim8_matching}
\end{table}
%
Lastly, in Table~\ref{Tab:real_triplet_dim8_matching}, we provide the complete tree-level matching results at dimension-eight. Here, for comparison, we focus exclusively on those operators whose coefficients were previously derived in Ref.~\cite{Corbett:2021eux} employing the field redefinition of the Higgs field. We can connect the structure of $\widetilde{\mathcal{O}}_{H^6\mathcal{D}^2,2}^{(8)}$ given in Ref.~\cite{Corbett:2021eux} with $\mathcal{O}_{H^6\mathcal{D}^2,1}^{(8)}$ and $\mathcal{O}_{H^6\mathcal{D}^2,2}^{(8)}$ (see appendix~\ref{app:op-structures} for the explicit structures of the operators), in the following way:
\begin{multline}
            (H^{\dagger}H)(H^{\dagger}\sigma^I H)(\mathcal{D}_{\mu}H^{\dagger}\sigma^I \mathcal{D}^{\mu}H)=2\,(H^{\dagger}H)(\mathcal{D}_{\mu}H^{\dagger}H)(H^{\dagger}\mathcal{D}^{\mu}H)-(H^{\dagger}H)^2(\mathcal{D}_{\mu}H^{\dagger}\mathcal{D}^{\mu}H).
\end{multline}
These results are in agreement with the expressions provided for the tree-level matching of the dimension-eight coefficients in Table 10 of Ref.~\cite{Corbett:2021eux}.
The remaining structures that arise after integrating out the heavy triplet scalar at dimension-eight, mainly at tree-level, including two and four-fermionic operators are shown in Table~\ref{Tab:additional_dim8_triplet_scalar}.
		\begin{table}[!t]
			\centering\scriptsize
			\renewcommand{\arraystretch}{2.6}
			\begin{tabular}{||c|c||c|c|c|c|}
				\hline
													
				$\mathcal{O}_{\psi^4H^2,1}^{(8)}$&
				\multicolumn{2}{c||}{$\frac{k^2}{m_{\Phi}^6}\,Y_{\text{SM}}^2$}&
				$\mathcal{O}_{\psi^4H^2,2}^{(8)}$&
				\multicolumn{2}{c|}{$\frac{k^2}{2m_{\Phi}^6}\,Y_{\text{SM}}^2$}\\
				\hline	
				
				$\mathcal{O}_{\psi^2H^3
					\mathcal{D}^2,1}^{(8)}$&
				$-\frac{2k^2}{m_{\Phi}^6}\,Y_{\text{SM}}$&
				$\mathcal{O}_{H^4\mathcal{D}^4,1}^{(8)}$&
				\multicolumn{1}{c||}{$\frac{4k^2}{m_{\Phi}^6}$}&
				$\mathcal{O}_{H^4\mathcal{D}^4,3}^{(8)}$&
				$-\frac{2k^2}{m_{\Phi}^6}$\\
				\hline				
				
			\end{tabular}
			\caption{\small Dimension-eight matching coefficients for the \textbf{real-triplet scalar extension} of SM, Eq.~\eqref{eq:real_triplet_Lagrangian}.}
			\label{Tab:additional_dim8_triplet_scalar}
		\end{table}
		
\section{Example models}
\label{sec:example}
This section applies the formalism described above to several example models to generate dimension-eight operators. 
We use \texttt{CoDEx}~\cite{Bakshi:2018ics} to obtain the operators and associated WCs in the SILH set at dimension six up to one loop, which we tabulate for each model. The coefficients are passed through Eqs.~\eqref{eq:SILH_to_mixed} to \eqref{eq:extra-term} that yield the contribution of dimension-six operators to dimension-eight operators. For clarity, we only present the leading contribution from the dimension-six tree-level generated operators and the direct integrated-out contribution at dimension eight. Subleading (yet non-negligible) corrections to the coefficients that arise from loop-generated operators  can be obtained accordingly\footnote{In Refs.~\cite{Chala:2021pll,Chala:2021wpj} some of the dimension-eight operators upto one-loop order for a few models have been computed.}, and the complete list of contributions can be obtained from a Mathematica notebook provided \href{https://github.com/shakeel-hep/D8-WCs}{\underline{here}}. The models (apart from the leptoquark one) we discuss below are chosen as they generate operators at the tree level (see, e.g., the discussion in Refs.~\cite{DasBakshi:2021xbl,Naskar:2022rpg}). We can classify the contributions to WCs into the following two categories:
\begin{itemize}
    \item \underline{Tree-level contributions}: In this category, we only consider the contribution from those WCs generated at the tree level at dimension six. When the EOM is applied, they contribute on a par with the tree-level generated dimension-eight operators. Their combined effects are then considered to be the leading order contributions at dimension eight. We will mainly focus on this type of contribution and tabulate results for each model.
    It should be noted that these operators also receive subleading loop-induced contributions. The complete expressions for the coefficients can be found in the \mathematicanb{}.
    \item \underline{Loop-induced and/or higher order contributions}: If the redundant dimension-six operators are generated at one-loop-level or the coefficients introduced through the application of the EOM appear with loop-level contributions or both, the WCs contain $(1/16\pi^2)$ or $(1/16\pi^2)^2$ suppressions, depending on interference between tree and loop parts.
    We will not list these types of contributions here, except for the leptoquark model for demonstration purposes. However, as mentioned above, the provided \mathematicanb{} contains all contributions, and interested readers are referred to the here documented results.
\end{itemize}

\subsection{Complex Triplet Scalar}
\label{subsec:complex_triplet_example}
The SM can be extended with an electroweak complex triplet scalar $(\Delta)$ to explain the generation of neutrino masses through type-II seesaw mechanism~\cite{Konetschny:1977bn,Cai:2017mow}. This model also offers interesting collider signatures comprising rare lepton number and flavour violating processes~\cite{Chakrabortty:2015zpm}. In addition to contemplating the phenomenological significance of this model, a consistent effort has been made in the recent past to explore the effective theory of such an extension, see Refs.~\cite{Anisha:2020ggj,Li:2022ipc,Du:2022vso}.
The BSM part of the Lagrangian reads:
		\begin{multline}\label{eq:ewrts_Lagrangian}
			\mathcal{L}_{\Delta} \supset (\mathcal{D}_{\mu}\Delta^{\dagger})(\mathcal{D}^{\mu}\Delta)-m_{\Delta}^2(\Delta^{\dagger}\Delta)-\big[\lambda_{\Delta}H^{\dagger}\sigma^I\widetilde{H}\Delta^{\dagger}+\text{h.c.}\big]-\lambda_1(\Delta^{\dagger}\Delta)^2\\
			-\lambda_2\,(\Delta^{\dagger} T^{I}\Delta)(\Delta^{\dagger} T^{I}\Delta)-\lambda_3(H^{\dagger}H)(\Delta^{\dagger}\Delta)-\lambda_4(H^{\dagger}\sigma^I H)(\Delta^{\dagger}T^{I}\Delta),
		\end{multline}		
(we neglect the interaction with the fermion fields in the following).
As the Lagrangian contains a linear interaction for the field $\Delta$, the renormalisable structure $(H^{\dagger}H)^2$ gets an extra contribution proportional to the coupling $\widetilde{\lambda} = 2\lambda_{\Delta}^2/m_{\Delta}^2$.

Table~\ref{Tab:complex_triplet_SILH_coefficients} contains the complete matching at one-loop-order for the dimension-six SILH coefficients. 
%
\begin{table}[!t]
\centering
\renewcommand{\arraystretch}{2.5} {\scriptsize
\begin{tabular}{|c|c|c|c|}
    \hline
    \textsf{\quad SILH Op.} &
    \textsf{\quad Wilson Coefficients}&
    \textsf{\quad SILH Op.}&
    \textsf{\quad Wilson Coefficients}\\
    \hline
		
    \multirow{5}{*}{$\mathcal{O}_{6}$} &
    $\frac{\lambda_4 \lambda_\Delta^2}{2m_\Delta^4}-\frac{2 \lambda_3\lambda_\Delta^2}{m_\Delta^4} -\biggl \{\frac{\lambda_3^3}{32 \pi^2 m_\Delta^2}-\frac{\lambda_3 \lambda_4^2}{64 \pi^2 m_\Delta^2}$ &
    \multirow{2}{*}{$\mathcal{O}_{H}$} &
    $\frac{2 \lambda_\Delta^2}{m_\Delta^4}+\biggl \{\frac{\lambda_3^2}{32\pi^2 m_\Delta^2} \biggr\} -\biggl [\frac{3\lambda_3 \lambda_\Delta^2}{4\pi^2 {m_\Delta^4}}+\frac{\lambda_4 \lambda_\Delta^2}{8 \pi^2 m_\Delta^4} $ \\
		
    &
    $-\frac{2 \lambda_1 \lambda_3 \lambda_\Delta^2}{\pi^2 m_\Delta^4}-\frac{\lambda_2 \lambda_3 \lambda_\Delta^2}{\pi^2 m_\Delta^4}-\frac{\lambda_1 \lambda_4 \lambda_\Delta^2}{2\pi^2 m_\Delta^4}-\frac{\lambda_2 \lambda_4 \lambda_\Delta^2}{4 \pi^2 m_\Delta^4} \biggr \}$ &
    &
    $+\frac{23 \lambda \lambda_\Delta^2}{8 \pi^2 m_\Delta^4}- \frac{11 \lambda_\Delta^4}{2\pi^2 m_\Delta^6} \biggr ]$ \\
    \cline{3-4}
  
    &
    $+\biggl [\frac{13 \lambda_3^2 \lambda_\Delta^2}{8 \pi^2 m_\Delta^4}-\frac{\lambda_3 \lambda_4 \lambda_\Delta^2}{2\pi^2 m_\Delta^4}-\frac{8 \lambda_3 \lambda \lambda_\Delta^2}{\pi^2
    m_\Delta^4}+\frac{3\lambda_4^2 \lambda_\Delta^2}{64 \pi ^2 m_\Delta^4}$&
    \multirow{3}{*}{$\mathcal{O}_{T}$}&
    $-\frac{2 \lambda_\Delta^2}{m_\Delta^4}-\biggl \{\frac{2 \lambda_1 \lambda_\Delta^2}{\pi^2 m_\Delta^4}- \frac{\lambda_2 \lambda_\Delta^2}{\pi^2 m_\Delta^4}$ \\

    &
   $-\frac{5 \lambda_1 \lambda_\Delta^4}{\pi^2 m_\Delta^6}-\frac{5\lambda_2 \lambda_\Delta^4}{\pi^2 m_\Delta^6}+\frac{24 \lambda_3\lambda_\Delta^4}{\pi^2 m_\Delta^6} +\frac{7 \lambda_4 \lambda \lambda \Delta^2}{4 \pi^2 m_\Delta^4}$&
   &
   $+\frac{\lambda_4^2}{768 \pi^2 m_\Delta^2}\biggr \}+\biggl [\frac{\lambda_3 \lambda_\Delta^2}{2 \pi^2 m_\Delta^4}-\frac{11
   \lambda_4 \lambda_\Delta^2}{24 \pi^2 m_\Delta^4}$
   \\
    
    &
    $+\frac{11 \lambda^2
   \lambda_\Delta^2}{\pi^2 m_\Delta^4}-\frac{41\lambda^4 \lambda_\Delta^4}{8 \pi^2 m_\Delta^6}-\frac{62
   \lambda \lambda_\Delta^4}{\pi^2 m_\Delta^6}+\frac{74\lambda_\Delta^6}{\pi ^2 m_\Delta^8} \biggr]$&
    &
    $-\frac{3 \lambda \lambda_\Delta^2}{8 \pi^2 m_\Delta^4}+\frac{4 \lambda_\Delta^4}{3 \pi^2 m_\Delta^6}\biggr]$ \\
    \hline

    \multirow{2}{*}{$\mathcal{O}_R$}&
    $\frac{4 \lambda_\Delta^2}{m_\Delta^4}+\biggl \{\frac{4 \lambda_1 \lambda_\Delta^2}{\pi^2 m_\Delta^4}+\frac{2 \lambda_2 \lambda_\Delta^2}{\pi^2 m_\Delta^4}+\frac{\lambda_4^2}{384 \pi^2 m_\Delta^2} \biggr\}$&
    $\mathcal{O}_{W}$&
    $-\biggl[\frac{\lambda_\Delta^2}{72 \pi^2 m_\Delta^4}\biggr]$ \\
    \cline{3-4}
		
    &
    $-\biggl[\frac{13\lambda_3 \lambda_\Delta^2}{8 \pi^2 m_\Delta^4}+\frac{\lambda_4 \lambda_\Delta^2}{6 \pi^2 m_\Delta^4}+ \frac{11\lambda \lambda_\Delta^2}{2 \pi^2 m_\Delta^4} -\frac{37\lambda_\Delta^4}{3\pi^2 m_\Delta^6}\biggr]$&
    $\mathcal{O}_{B}$ &
    $\biggl[\frac{11 \lambda_\Delta^2}{24 \pi^2 m_\Delta^4}\biggr]$ \\
    \hline	

    $\mathcal{O}_{D}$ &
    $\biggl[\frac{\lambda_\Delta^2}{8 \pi^2 m_\Delta^4}\biggr]$&
    $\mathcal{O}_{WW}$ &
    $\biggl\{\frac{\lambda_3}{96 \pi^2 m_\Delta^2}\biggr\}+\biggl[\frac{25 \lambda_\Delta^2}{192 \pi^2 m_\Delta^4}\biggr]$ \\
    \hline
     
    $ \mathcal{O}_{WB}$ &
    $\biggl\{\frac{\lambda_4}{384\pi^2 m_\Delta^2}\biggr\}-\biggl[\frac{13 \lambda_\Delta^2}{96 \pi^2 m_\Delta^4}\biggr]$ &
    $\mathcal{O}_{BB}$&
    $\biggl\{\frac{\lambda_3}{64 \pi^2 m_\Delta^2}\biggr\}+\biggl[\frac{11 \lambda_\Delta^2}{64 \pi^2 m_\Delta^4}\biggr]$ \\
    \hline
        
    $\mathcal{O}_{2W}$&
    $\biggl\{\frac{g_W^2}{240 \pi^2 m_\Delta^2}\biggr\}$&
    $\mathcal{O}_{2B}$&
    $\biggl\{\frac{g_Y^2}{160 \pi^2 m_\Delta^2}\biggr\}$\\
    \hline
    
\end{tabular}}
\caption{\small WCs of dimension-six SMEFT operators in the SILH set after integrating out the \textbf{Complex Triplet Scalar} Eq.~\eqref{eq:ewrts_Lagrangian}. The terms within braces $(\{\})$ denote the contribution from pure heavy loops, whereas the brackets $(\left[ \,\right])$ mark the contribution from light-heavy mixed loops. We only use the uncoloured coefficients for further calculation here. The complete calculation can be found in the provided \mathematicanb{}.}
\label{Tab:complex_triplet_SILH_coefficients}
\end{table}		
After the tree-level integrating we obtain the following WCs at dimension-eight:
      \begin{eqnarray}\label{eq:complex-triplet-dim8-coefficients}
          \zeta_1^{(8)}&=&\big(\frac{2(\lambda_3-\lambda_4)^2\lambda_{\Delta}^2}{m_{\Delta}^6}-\frac{8(\lambda_1+\lambda_2)\lambda_{\Delta}^4}{m_{\Delta}^8}\big);
          \hspace{3mm}\zeta_2^{(8)}=\frac{8(\lambda_3-\lambda_4)\lambda_{\Delta}^2}{m_{\Delta}^6};\hspace{3mm}\zeta_3^{(8)}=\frac{4(\lambda_3-\lambda_4)\lambda_{\Delta}^2}{m_{\Delta}^6};\nonumber\\
          \zeta_4^{(8)}&=&\frac{8\lambda_{\Delta}^2}{m_{\Delta}^6};\hspace{3mm}\xi_1^{(8)}=-\frac{2(\lambda_3-\lambda_4)\lambda_{\Delta}^2}{m_{\Delta}^6}; \hspace{4mm}\xi_3^{(8)}=\frac{8\lambda_{\Delta}^2}{m_{\Delta}^6};\hspace{4mm}
          \xi_6^{(8)}=\frac{4\lambda_{\Delta}^2}{m_{\Delta}^6};\hspace{4mm}\xi_7^{(8)}=\frac{4\lambda_{\Delta}^2}{m_{\Delta}^6}.
      \end{eqnarray}
As mentioned before, the total contribution to the dimension-eight operators is categorised into two categories depending on how they contribute. Below we write down the operators in their respective categories.
\begin{itemize}
    \item \underline{Tree-level contribution}: $\mathcal{O}_{H^6\mathcal{D}^2,1}^{(8)}, \,\mathcal{O}_{H^6\mathcal{D}^2,2}^{(8)},\,\mathcal{O}_{H}^{(8)},\,\mathcal{O}_{\psi^2 H^5}^{(8)},\,\mathcal{O}_{\psi^2H^3\mathcal{D}^2,1}^{(8)}, \,\mathcal{O}_{\psi^4H^2,1}^{(8)}$. The WCs corresponding to these operators are listed in Table~\ref{Tab:complex-triplet_dimension_8_coefficients}.
    \item \underline{Loop-induced and/or higher order contribution}: $ \mathcal{O}_{\psi^2H^4\mathcal{D},1}^{(8)},\, \mathcal{O}_{\psi^2H^4\mathcal{D},2}^{(8)},\,\mathcal{O}_{\psi^4\mathcal{D}H,1}^{(8)},\\\mathcal{O}_{\psi^4\mathcal{D}H,2}^{(8)},\,\mathcal{O}_{\psi^2H^3\mathcal{D}^2,2}^{(8)},\,\mathcal{O}_{H^4X^2}^{(8)},\mathcal{O}_{\psi^2X^2H}^{(8)},\mathcal{O}_{\psi^4H^2,2}^{(8)}$. The WCs corresponding to these operators are listed in the \mathematicanb{}.
\end{itemize}
      
\begin{table}[!t]
\centering\scriptsize
\renewcommand{\arraystretch}{2.6}
    \begin{tabular}{|c|c|c|c|}
    \hline
    {\textsf{Operator}} &
    {\textsf{Wilson coefficients}} &
    {\textsf{Operator}} &
    {\textsf{Wilson coefficients}} \\
    \hline

    $\mathcal{O}_{H^6\mathcal{D}^2,1}^{(8)}$&
    $\frac{8(\lambda_3-\lambda_4)\lambda_{\Delta}^2}{m_{\Delta}^6}-\frac{20\lambda_{\Delta}^4}{m_{\Delta}^8}$&
    $\mathcal{O}_{H^6\mathcal{D}^2,2}^{(8)}$
    &$-\frac{8(4\lambda-\lambda_3+\lambda_4)\lambda_{\Delta}^2}{m_{\Delta}^6}+\frac{32\lambda_{\Delta}^4}{m_{\Delta}^8}$ \\
    \hline
    
    \multirow{2}{*}{$\mathcal{O}_{H}^{(8)}$} &
    $\frac{2\big(16\lambda^2+4\lambda(\lambda_3-\lambda_4)+(\lambda_3-\lambda_4)^2\big)\lambda_{\Delta}^2}{m_{\Delta}^6}$ &
    $\mathcal{O}_{\psi^2 H^5}^{(8)}$ &
    $\bigg(\frac{2(8\lambda+\lambda_3-\lambda_4)\lambda_{\Delta}^2}{m_{\Delta}^6}-\frac{36\lambda_{\Delta}^4}{m_{\Delta}^8}\bigg)\,Y_{\text{SM}}$ \\
    \cline{3-4}

    &
    $-\frac{8(21\lambda+\lambda_1+\lambda_2-\lambda_3+\lambda_4)\lambda_{\Delta}^4}{m_{\Delta}^8}+\frac{208\lambda_{\Delta}^6}{m_{\Delta}^{10}}$&
    $\mathcal{O}_{H^4\mathcal{D}^4,1}^{(8)}$&
    $\frac{8\lambda_{\Delta}^2}{m_{\Delta}^6}$
    \\
    \hline
				
    $\mathcal{O}_{\psi^2H^3\mathcal{D}^2,1}^{(8)}$ &
    $-\frac{8\lambda_{\Delta}^2}{m_{\Delta}^6}\,Y_{\text{SM}}$&
    $\mathcal{O}_{\psi^4H^2,1}^{(8)}$&
    $\frac{8\lambda_{\Delta}^2}{m_{\Delta}^6}\,Y_{\text{SM}}^2$\\
    \hline	

    \end{tabular}
    \caption{\small Total dimension-eight tree-level contribution after integrating out the \textbf{Complex Triplet Scalar} of Eq.~\eqref{eq:ewrts_Lagrangian}. Complete results at one-loop with additional operators are presented in the \mathematicanb{}.}
    \label{Tab:complex-triplet_dimension_8_coefficients}
\end{table}

\subsection{General Two Higgs Doublet Model}
\label{subsec:2HDM_example}
One of the simplest and 
well-motivated extensions of the SM Higgs sector is the inclusion of an additional $SU(2)_{\text{L}}$ scalar doublet ($\mathcal{H}$) with hypercharge $\mathrm{Y}=1/2$, the well-known two-Higgs doublet model (2HDM) \cite{Branco:2011iw,KIM19871}. Many UV complete theories contain a 2HDM in their minimal versions. This model has also been well discussed within SMEFT framework by integrating out the additional heavy Higgs doublet leading to dimension-six effective operators at one-loop~\cite{deBlas:2017xtg,DasBakshi:2020pbf,Anisha:2021hgc}. The relevant part of the BSM Lagrangian reads
\begin{multline}\label{eq:2HDM_Lagrangian}
			\mathcal{L}_{\mathcal{H}} \supset (\mathcal{D}_{\mu}\mathcal{H}^{\dagger})(\mathcal{D}^{\mu}\mathcal{H})-m_{\mathcal{H}}^2\mathcal{H}^{\dagger} \mathcal{H}-\frac{\lambda_{\mathcal{H}}}{4}(\mathcal{H}^{\dagger} \mathcal{H})^2+\big(\eta_H (\widetilde{H}^{\dagger}\widetilde{H})+\eta_{\mathcal{H}}(\mathcal{H}^{\dagger}\mathcal{H})\big)\big(\widetilde{H}^{\dagger}\mathcal{H}+\mathcal{H}^{\dagger}\widetilde{H}\big)\\
			-\lambda_1 \big(\widetilde{H}^{\dagger}\widetilde{H}\big)(\mathcal{H}^{\dagger}\mathcal{H})^2-\lambda_2 (\mathcal{H}^{\dagger}\widetilde{H})(\widetilde{H}^{\dagger}\mathcal{H})-\lambda_3 \big[(\widetilde{H}^{\dagger}\mathcal{H})^2+(\mathcal{H}^{\dagger}\widetilde{H})^2\big].
\end{multline}	
The Wilson coefficients of dimension-six operators in the SILH set are presented in Table~\ref{Tab:2HDM_SILH_coefficients}.
  	\begin{table}[!t]
	\centering
	\renewcommand{\arraystretch}{2.6} \scriptsize
\resizebox{\textwidth}{!}{\begin{tabular}{|c|c|c|c|}
		\hline
		\textsf{\quad SILH Op.}&
		\textsf{\quad Wilson Coefficients}&
		\textsf{\quad SILH Op.}&
		\textsf{\quad Wilson Coefficients}\\
		\hline
		
	    \multirow{3}{*}{$\mathcal{O}_{6}$}&
		$\frac{\eta_H^2}{m_\mathcal{H}^2}+\biggl \{\frac{3\eta_H^2\lambda_{\mathcal{H}}}{32m_\mathcal{H}^2\pi^2}+\frac{3\eta_H \eta_{\mathcal{H}}\lambda_{2}}{8m_\mathcal{H}^2\pi^2}-\frac{\lambda_1^3}{48m_{\mathcal{H}}^2\pi^2}$&\multirow{2}{*}{$\mathcal{O}_{H}$}
		&$\biggl\{-\frac{3 \eta_H \eta_\mathcal{H}}{8 \pi^2
   m_\mathcal{H}^2}+\frac{\lambda_1^2}{48 \pi ^2
   m_\mathcal{H}^2}+\frac{\lambda_1 \lambda_2}{48
   \pi ^2 m_\mathcal{H}^2}$
        \\
        
        &
        $+\frac{3\eta_H\eta_{\mathcal{H}}\lambda_{2}}{8m_{\mathcal{H}}^2\pi^2}-\frac{\lambda_1^2\lambda_2}{32m_{\mathcal{H}}^2\pi^2}-\frac{\lambda_2^3}{96m_{\mathcal{H}}^2\pi^2}-\frac{\lambda_1\lambda_3^2}{8m_{\mathcal{H}}^2\pi^2}$&
        &
        $+\frac{\lambda_2^2}{192 \pi ^2
   m_\mathcal{H}^2}+\frac{\lambda_3^2}{48 \pi ^2
   m_\mathcal{H}^2}\biggr\}+\biggl[\frac{5\eta_H^2}{16 \pi ^2 m_\mathcal{H}^2}\biggr]$
        \\
        \cline{3-4}
        &
        $-\frac{\lambda_2\lambda_3^2}{8m_{\mathcal{H}}^2\pi^2}\biggr\}+\biggl[\frac{15\eta_{H}^2\lambda}{8m_{\mathcal{H}}^2\pi^2}-\frac{3\eta_H^2\lambda_1}{4m_{\mathcal{H}}^2\pi^2}-\frac{13\eta_H^2\lambda_2}{16\mathcal{H}^2\pi^2}-\frac{7\eta_H^2\lambda_3}{4\mathcal{H}^2\pi^2}\biggr]$&
        $\mathcal{O}_{2B}$&
        $\biggl\{\frac{g_{_W}^2}{960m_{\mathcal{H}}^2\pi^2}\biggr\}$
        \\
		\hline

		$\mathcal{O}_R$&
		$-\biggl\{\frac{3\eta_H\eta_{\mathcal{H}}}{8m_{\mathcal{H}}^2\pi^2}+\frac{\lambda_2^2}{96m_{\mathcal{H}}^2\pi^2}+\frac{\lambda_3^2}{24m_{\mathcal{H}}^2\pi^2}\biggr\}+\biggl[\frac{\eta_H^2}{8m_{\mathcal{H}}^2\pi^2}\biggr]$&
		$\mathcal{O}_{T}$& 
		$\biggl\{\frac{\lambda_2^2}{192m_{\mathcal{H}}^2\pi^2}-\frac{\lambda_3^2}{48m_{\mathcal{H}}^2\pi^2}\biggr\}$\\
		\hline

        $ \mathcal{O}_{WW}$&
        $\biggl\{\frac{\lambda_1}{384m_{\mathcal{H}}^2\pi^2}+\frac{\lambda_2}{768m_{\mathcal{H}}^2\pi^2}\biggr\}$&
        $\mathcal{O}_{BB}$&
        $\biggl\{\frac{\lambda_1}{384m_{\mathcal{H}}^2\pi^2}+\frac{\lambda_2}{768m_{\mathcal{H}}^2\pi^2}\biggr\}$\\
        \hline

        $\mathcal{O}_{WB}$&
        $\biggl\{\frac{\lambda_2}{384m_{\mathcal{H}}^2\pi^2}\biggr\}$&
        $\mathcal{O}_{2W}$&
        $\biggl\{\frac{g_{_W}^2}{960m_{\mathcal{H}}^2\pi^2}\biggr\}$\\
        \hline

	\end{tabular}}
	\caption{\small WCs of dimension-six SMEFT operators in the SILH set after integrating out an additional \textbf{Higgs Doublet}, Eq.~\eqref{eq:2HDM_Lagrangian}. The terms within braces $(\{\})$ denote the contribution from pure heavy loops, whereas the brackets $(\left[ \,\right])$ mark the contribution from light-heavy mixed loops. We only use the uncoloured coefficients for further calculation here. The complete calculation can be found in the \mathematicanb{}.}
	\label{Tab:2HDM_SILH_coefficients}
\end{table}
After integrating out at tree-level the non-zero dimension-eight coefficients are given by
      \begin{eqnarray}\label{eq:2HDM-dim8-coefficients}
          \zeta_1^{(8)}=-\frac{\eta_H^2}{m_{\mathcal{H}}^4}(\lambda_1+\lambda_2+2\lambda_3); \hspace{4mm}\zeta_3^{(8)}=-\frac{\eta_H^2}{m_{\mathcal{H}}^4}; \hspace{4mm}\xi_1^{(8)}=-\frac{\eta_{H}^2}{m_{\mathcal{H}}^4}.
      \end{eqnarray}
We split the operators into our categories:
\begin{itemize}
    \item \underline{Tree-level contribution}: $\mathcal{O}_{H^6\mathcal{D}^2,1}^{(8)},\,  \mathcal{O}_{H}^{(8)},\,\mathcal{O}_{\psi^2 H^5}^{(8)} $. The WCs corresponding to these operators are listed in Table~\ref{Tab:2HDM-d8}.
    \item \underline{Loop-induced and/or higher order contribution}: $ \mathcal{O}_{H^6\mathcal{D}^2,2}^{(8)},\,\mathcal{O}_{\psi^2H^4\mathcal{D},1}^{(8)},\, \mathcal{O}_{\psi^2H^4\mathcal{D},2}^{(8)},\\\mathcal{O}_{\psi^4\mathcal{D}H,1}^{(8)},$
    $\mathcal{O}_{\psi^4\mathcal{D}H,2}^{(8)},\,\mathcal{O}_{\psi^2H^3\mathcal{D}^2,1}^{(8)},\,\mathcal{O}_{\psi^2H^3\mathcal{D}^2,2}^{(8)},\,\mathcal{O}_{H^4 X^2}^{(8)},\,\mathcal{O}_{\psi^2X^2H}^{(8)},\,\mathcal{O}_{\psi^4H^2,1}^{(8)},\,\mathcal{O}_{\psi^4H^2,2}^{(8)}$. 
    
    The WCs corresponding to these operators are provided in the \mathematicanb{}.
\end{itemize}		

\begin{table}[!t]
    \centering\scriptsize
    \renewcommand{\arraystretch}{2.6}
        \begin{tabular}{||c|c||c|c||c|c||}
	\hline
	\textsf{Operator} &
	\textsf{Wilson coefficients}&
        \textsf{Operator} &
	\textsf{Wilson coefficients} &
	\textsf{Operator} &
	\textsf{Wilson coefficients} \\
	\hline

	$\mathcal{O}_{H}^{(8)}$&
	$\frac{\eta_H^2(4\lambda-\lambda_1-\lambda_2-2\lambda_3)}{m_{\mathcal{H}}^4}$&
	$\mathcal{O}_{\psi^2 H^5}^{(8)}$&
	$\frac{\eta_H^2}{m_{\mathcal{H}}^4}\,Y_{\text{SM}}$&
        $\mathcal{O}_{H^6\mathcal{D}^2,1}^{(8)}$&
	$-\frac{\eta_H^2}{m_{\mathcal{H}}^4}$ \\
	\hline
	
        \end{tabular}
	\caption{\small Total dimension-eight tree-level contribution after integrating out additional an \textbf{Higgs Doublet}, Eq.~\eqref{eq:2HDM_Lagrangian}. Complete contributions at one-loop including additional operators are presented in the \mathematicanb{}. }
	\label{Tab:2HDM-d8}
\end{table}		

\subsection{Complex quartet Scalar (hypercharge $\mathrm{Y}=3/2)$ }
To generate neutrino masses through higher dimensional operators, an $SU(2)_{\text{L}}$ quartet ($\Sigma$) with hypercharge $\mathrm{Y}=3/2$ can be added to the SM \cite{Nomura:2018cle,Babu:2009aq,Bambhaniya:2013yca}. Focusing on this part of the BSM Lagrangian,
  \begin{eqnarray}
  \label{eq:quartet}
      \mathcal{L}_{\Sigma}&\supset&(\mathcal{D}_\mu\Sigma^\dagger)(\mathcal{D}^\mu\Sigma)-M_{\Sigma}^2\Sigma^\dagger\Sigma+(\eta_{\Sigma}\; \Sigma^\dagger_{jkl}H^j H^k H^l + \text{h.c.})-k_{\Sigma_1} (H^\dagger H)(\Sigma^\dagger \Sigma) \nonumber\\ 
      & & \hspace*{2.5cm} - k_{\Sigma_2} (H_m^\dagger H^n)(\Sigma^\dagger_{jkn} \Sigma^{jkm}) - \lambda_{\Sigma_1}(\Sigma^\dagger \Sigma)^2 -\lambda_{\Sigma_2}(\Sigma^\dagger T^a \Sigma)^2 \,,
  \end{eqnarray}
we can integrate out the heavy scalar and match to the dimension-six SMEFT operators~(see also~\cite{Murphy:2020rsh, Anisha:2021hgc,DasBakshi:2020pbf}).
After matching we obtain the effective operators and associated WCs in terms of the UV parameters, shown in Table~\ref{Tab:quartet_SILH_coefficients}. 
\begin{table}[!b]
    \centering
    \renewcommand{\arraystretch}{2.6} \scriptsize
 \resizebox{\textwidth}{!}{\begin{tabular}{|c|c|c|c|}
	\hline
	\textsf{ SILH Op.} &
	\textsf{ Wilson Coefficients} &
	\textsf{ SILH Op.} &
	\textsf{ Wilson Coefficients} \\
	\hline
		
	\multirow{2}{*}{$\mathcal{O}_{6}$}&
	$\frac{\eta_{\Sigma}^2}{M_{\Sigma}^2} -\biggl\{\frac{k_{\Sigma_1}^2 k_{\Sigma_2}}{16 \pi ^2       M_{\Sigma}^2}-\frac{k_{\Sigma_1}^3}{24 \pi^2 M_\Sigma^2}-\frac{7 k_{\Sigma_1} k_{\Sigma_2}^2}{144         \pi^2 M_\Sigma^2}\biggr\} -\bigg[\frac{9 \eta_{\Sigma}^2 k_{\Sigma_1}}{8 \pi^2 M_\Sigma^2}\bigg]$ &
        $\mathcal{O}_{H}$ &
	$\biggl\{\frac{k_{\Sigma_1}^2}{24 \pi^2 M_\Sigma^2}+\frac{k_{\Sigma_1} k_{\Sigma_2}}{24 \pi^2     M_\Sigma^2}+\frac{k_{\Sigma_2}^2}{96 \pi^2 M_\Sigma^2}\biggr\} + \biggl[\frac{3 \eta_{\Sigma}^2}{8    \pi^2 M_\Sigma^2}\biggr]$ \\
        \cline{3-4}        
        
        & 
        $-\biggl\{\frac{k_{\Sigma_2}^3}{72 \pi^2 M_\Sigma^2}+\frac{5 \eta_{\Sigma}^2 \lambda_{\Sigma_1}}{8 \pi ^2 M_\Sigma^2}+\frac{15 \eta_{\Sigma}^2 \lambda_{\Sigma_2}}{32 \pi ^2 M_\Sigma^2}\biggr\}-\bigg[\frac{19 \eta_{\Sigma}^2 k_{\Sigma_2}}{16 \pi ^2 M_\Sigma^2}\bigg] $&
        $\mathcal{O}_{2B}$&
        $\biggl\{\frac{3 g_Y^2}{160 \pi^2 M_\Sigma^2}\biggr\}$ \\
        \hline
        
	$\mathcal{O}_R$&
	$\biggl\{\frac{5 k_{\Sigma_2}^2}{432 \pi^2 M_\Sigma^2}\biggr\} + \biggl[\frac{3       \eta_{\Sigma} ^2}{4 \pi^2 M_\Sigma^2}\biggr]$&
	$\mathcal{O}_{T}$& 
	$\biggl\{\frac{5 k_{\Sigma_2}^2}{864 \pi ^2 M_\Sigma^2}\biggr\}-\biggl[\frac{3 \eta_{\Sigma}^2}{8\pi^2 M_\Sigma^2}\biggr]$\\
	\hline	

        $ \mathcal{O}_{WW}$&
        $\biggl\{\frac{5 k_{\Sigma_1}}{192 \pi^2 M_\Sigma^2}+\frac{5 k_{\Sigma_2}}{384 \pi^2 M_\Sigma^2}\biggr\}$&
        $\mathcal{O}_{BB}$&
        $\biggl\{\frac{3 k_{\Sigma_1}}{64 \pi^2 M_\Sigma^2}+\frac{3 k_{\Sigma_2}}{128 \pi ^2 M_\Sigma^2}\biggr\}$\\
        \hline
               
        $\mathcal{O}_{WB}$&
        $\biggl\{\frac{5 k_{\Sigma_2}}{192 \pi^2 M_\Sigma^2}\biggr\}$&
        $\mathcal{O}_{2W}$&
        $\biggl\{\frac{g_W^2}{96 \pi^2 M_\Sigma^2}\biggr\}$\\
        \hline        
	\end{tabular}}
	\caption{\small WCs of dimension-six SMEFT operators in the SILH set after integrating out the \textbf{Quartet Scalar} of Eq.~\eqref{eq:quartet}. The terms within braces $(\{\})$ denote the contribution from pure heavy loops, whereas the brackets $(\left[ \,\right])$ mark the contribution from light-heavy mixed loops. We only use the uncoloured coefficients for further calculation here. The complete calculation can be found in the \mathematicanb{}.}
	\label{Tab:quartet_SILH_coefficients}
\end{table}
Tree-level matching generates the following dimension-eight operators coefficients:
  \begin{eqnarray}
      \zeta_1^{(8)} =  -\frac{|\eta_{\Sigma}|^2 (k_{\Sigma_1}+k_{\Sigma_2})}{M_\Sigma^4};\; \zeta_2^{(8)}= -\frac{6|\eta_{\Sigma}|^2 }{M_\Sigma^4};\; \zeta_3^{(8)}= -\frac{6|\eta_{\Sigma}|^2 }{M_\Sigma^4}.
  \end{eqnarray}
Again, we split the operators into the two categories:
\begin{itemize}
    \item \underline{Tree-level contribution}: $\mathcal{O}_{H^6\mathcal{D}^2,1}^{(8)},\,  \mathcal{O}_{H}^{(8)},\,\mathcal{O}_{H^6\mathcal{D}^2,2}^{(8)}$. The WCs corresponding to these operators are listed in Table~\ref{Tab:complex-quartet_dimension_8_coefficients}.
    \item \underline{Loop-induced and/or higher order contribution}: $ \mathcal{O}_{\psi^2H^5}^{(8)},\,\mathcal{O}_{\psi^2H^4\mathcal{D},1}^{(8)},\, \mathcal{O}_{\psi^2H^4\mathcal{D},2}^{(8)},\,\mathcal{O}_{\psi^4\mathcal{D}H,1}^{(8)},$\\
    $\mathcal{O}_{\psi^4\mathcal{D}H,2}^{(8)},\,\mathcal{O}_{\psi^2H^3\mathcal{D}^2,1}^{(8)},\,\mathcal{O}_{\psi^2H^3\mathcal{D}^2,2}^{(8)},\,\mathcal{O}_{H^4 X^2}^{(8)},\,\mathcal{O}_{\psi^2X^2H}^{(8)},\,\mathcal{O}_{\psi^4H^2,1}^{(8)},\,\mathcal{O}_{\psi^4H^2,2}^{(8)}$. The WCs corresponding to these operators are shown in the \mathematicanb{}.
\end{itemize}

\begin{table}[!t]
    \centering\scriptsize
    \renewcommand{\arraystretch}{2.6}
        \begin{tabular}{||c|c||c|c||c|c||}
	\hline
	\textsf{Operator} &
	\textsf{Wilson coefficients} &
	\textsf{Operator} &
	\textsf{Wilson coefficients} &
        \textsf{Operator} &
	\textsf{ Wilson coefficients}\\
	\hline

	$\mathcal{O}_{H^6\mathcal{D}^2,1}^{(8)}$ &
	$-\frac{6 \eta_\Sigma^2}{m_\Sigma^4}$ &
	$\mathcal{O}_{H^6\mathcal{D}^2,2}^{(8)}$ &
	$-\frac{6 \eta_\Sigma^2}{m_\Sigma^4}$ &
        $\mathcal{O}_{H}^{(8)}$ &
	$-\frac{\eta_\Sigma^2 k_{\Sigma1}}{m_\Sigma^4}-\frac{\eta_\Sigma^2 k_{\Sigma2}}{m_\Sigma^4}$ \\
	\hline

        \end{tabular}
	\caption{\small Total dimension-eight tree-level contribution after integrating out the \textbf{Quartet Scalar} of Eq.~\eqref{eq:quartet}. Complete one-loop results including additional operators are available from the \mathematicanb{}. }
	\label{Tab:complex-quartet_dimension_8_coefficients}
    \end{table}		

\subsection{Real Singlet Scalar Model}
\label{subsec:real_singlet_example}
The addition of a real singlet scalar to the SM is motivated by a range of SM shortcomings, related to dark matter, baryogenesis, and the electroweak hierarchy problem \cite{McDonald:1993ex,Enqvist:2014zqa,Guo:2010hq}. This model has been discussed extensively within the EFT framework through a complete one-loop matching to the SMEFT up to dimension-six~\cite{Jiang:2018pbd,Haisch:2020ahr,DasBakshi:2020pbf,Anisha:2020ggj,Anisha:2021hgc,Dittmaier:2021fls}. Here, we systematically extend these results.
The Lagrangian involving the Real Singlet Scalar field $(\mathcal{S})$ is given by:
\begin{eqnarray}\label{eq:real_singlet_Lagrangian}
	\mathcal{L}_{\mathcal{S}} \supset \frac{1}{2} (\partial_{\mu}\mathcal{S})^2-\frac{1}{2}M_{\mathcal{S}}^2\mathcal{S}^2-\eta_{\mathcal{S}}(H^{\dagger}H)\mathcal{S}-k_{\mathcal{S}}(H^{\dagger}H)\mathcal{S}^2-\frac{1}{4!}\lambda_{\mathcal{S}}\mathcal{S}^4.
\end{eqnarray}
After integrating out the heavy field $\mathcal{S}$, we obtain an additional contribution to the quartic coupling of Higgs: $ \tilde{\lambda}=-\eta_{\mathcal{S}}^2/M_{\mathcal{S}}^2 $, along with the dimension-six SILH set operators as shown in Table~\ref{Tab:real_singlet_SILH_coefficients}.
\begin{table}[!t]
	\centering
	\renewcommand{\arraystretch}{2.6} \scriptsize
	\begin{tabular}{||c|c||c|c||c|c||c|c||}
		\hline
		
	\multirow{3}{*}{$\mathcal{O}_{6}$}&
	   \multicolumn{3}{c||}{$-\frac{\eta_\mathcal{S}^2 k_\mathcal{S}}{M_\mathcal{S}^4} -       \biggl\{\frac{\eta_\mathcal{S}^2 k_\mathcal{S} \lambda_\mathcal{S}}{16 \pi ^2 M_\mathcal{S}^4}-\frac{k_\mathcal{S}^3}{24 \pi ^2 M_\mathcal{S}^2}\biggr\}$}&
	   \multirow{2}{*}{$\mathcal{O}_{H}$} &
	   \multicolumn{3}{c||}{$\frac{\eta_\mathcal{S}^2}{M_\mathcal{S}^4} +\biggl\{ \frac{\eta_\mathcal{S}^2 \lambda_\mathcal{S}}{16 \pi ^2 M_\mathcal{S}^4}+\frac{k_\mathcal{S}^2}{48 \pi ^2 M_\mathcal{S}^2}\biggr\}$}
        \\
        
        &
        \multicolumn{3}{c||}{$+\biggl[\frac{11 \eta_\mathcal{S}^2 k_\mathcal{S}^2}{8 \pi^2 M_\mathcal{S}^4}+\frac{37 \eta_\mathcal{S}^4 k_\mathcal{S}}{16 \pi^2 M_\mathcal{S}^6}-\frac{3 \eta_\mathcal{S}^4 \lambda} {2 \pi^2 M_\mathcal{S}^6}+\frac{43 \eta_\mathcal{S}^6}{48 \pi^2 M_\mathcal{S}^8}$}&
        &
        \multicolumn{3}{c||}{$-\biggl[\frac{17 \eta\mathcal{S}^2 k_\mathcal{S}}{24 \pi^2 M_\mathcal{S}^4}+\frac{9 \eta_\mathcal{S}^2 \lambda}{32 \pi^2 M_\mathcal{S}^4}-\frac{5 \eta_\mathcal{S}^4}{12 \pi^2 M_\mathcal{S}^6}\biggr]$}
        \\
        \cline{5-8}
        
        &
        \multicolumn{3}{c||}{$-\frac{3 \eta_\mathcal{S}^2 k_\mathcal{S} \lambda}{2 \pi^2 M_\mathcal{S}^4}+\frac{9 \eta_\mathcal{S}^2 \lambda^2}{16 \pi^2 M_\mathcal{S}^4}-\frac{\eta_\mathcal{S}^4 \lambda_\mathcal{S}}{32 \pi^2 M_\mathcal{S}^6}\biggr]$}&
        $\mathcal{O}_{W}$&
        $\biggl[-\frac{7 \eta_\mathcal{S}^2}{288 \pi^2 M_\mathcal{S}^4}\biggr]$&
        $\mathcal{O}_{B}$&
        $\biggl[-\frac{7 \eta_\mathcal{S}^2}{288 \pi^2 M_\mathcal{S}^4}\biggr]$
        \\
	\hline

	$\mathcal{O}_D$&
    $\biggl[\frac{\eta_\mathcal{S}^2}{96 \pi^2 M_\mathcal{S}^4}\biggr]$&
        $ \mathcal{O}_{WB}$&
      $\biggl[\frac{\eta_\mathcal{S}^2}{128 \pi^2 M_\mathcal{S}^4}\biggr]$&
	$\mathcal{O}_{WW}$& 
	$\biggl[\frac{\eta_\mathcal{S}^2}{256 \pi^2 M_\mathcal{S}^4}\biggr]$&
        $\mathcal{O}_{BB}$&
	$\biggl[\frac{\eta_\mathcal{S}^2}{256 \pi^2 M_\mathcal{S}^4}\biggr]$\\
	\hline	

        \hline
        
	\end{tabular}
	\caption{\small WCs of dimension-six SMEFT operators in the SILH set after integrating out the \textbf{Real Singlet Scalar} of Eq.~\eqref{eq:real_singlet_Lagrangian}. The terms within braces $(\{\})$ denote the contribution from pure heavy loops, whereas the brackets $(\left[ \,\right])$ mark the contribution from light-heavy mixed loops.}
	\label{Tab:real_singlet_SILH_coefficients}
\end{table}
Since no redundant operator at dimension-six is generated at tree-level, there is no tree-level contribution to dimension-eight operators from dimension-six. Thus the dominant contribution arises solely from removing redundancies at dimension-eight level itself. The non-zero coefficients of dimension-eight operators generated via integrating out are
\begin{eqnarray}
    \zeta_1^{(8)} &=& \frac{2 \eta_{\mathcal{S}}^2 k_{\mathcal{S}}^2}{M_\mathcal{S}^6}-\frac{\lambda_{\mathcal{S}}\eta_{\mathcal{S}}^4}{24 M_\mathcal{S}^8};  \hspace{2mm}\zeta_3^{(8)}=\frac{4\eta_{\mathcal{S}}^2 k_{\mathcal{S}}}{M_{\mathcal{S}}^6};\hspace{2mm}\zeta_6^{(8)}=\frac{2\eta_{\mathcal{S}}^2 }{M_{\mathcal{S}}^6};\;  \nonumber\\ 
    \xi_1^{(8)}&=&\frac{2\eta_{\mathcal{S}}^2 k_{\mathcal{S}}}{M_{\mathcal{S}}^6};\; \xi_2^{(8)}=\frac{2\eta_{\mathcal{S}}^2 k_{\mathcal{S}}}{M_{\mathcal{S}}^6};\hspace{2mm}\xi_4^{(8)}=\frac{\eta_{\mathcal{S}}^2 k_{\mathcal{S}}}{2 M_{\mathcal{S}}^6};\hspace{2mm} \xi_7^{(8)}=\frac{\eta_{\mathcal{S}}^2 k_{\mathcal{S}}}{M_{\mathcal{S}}^6}.
\end{eqnarray}  
\begin{table}[!t]
    \centering\scriptsize
    \renewcommand{\arraystretch}{2.6}
        \begin{tabular}{||c|c||c|c|c|c||}
	\hline
					
	\multirow{3}{*}{$\mathcal{O}_{H}^{(8)}$}&
        \multicolumn{2}{c||}{$\frac{2\eta_{\mathcal{S}}^2k_{\mathcal{S}}^2}{m_{\mathcal{S}}^6}-\frac{8\eta_{\mathcal{S}}^2k_{\mathcal{S}}\lambda}{m_{\mathcal{S}}^6}$}&
	$\mathcal{O}_{H^6\mathcal{D}^2,1}^{(8)}$ &
        \multicolumn{2}{c||}{$\frac{4\eta_{\mathcal{S}}^2k_{\mathcal{S}}}{m_{\mathcal{S}}^6}-\frac{8\lambda\,\eta_{\mathcal{S}}^2k_{\mathcal{S}}}{m_{\mathcal{S}}^6}$} \\
        \cline{4-6}
        
        &
        \multicolumn{2}{c||}{$+\frac{8\eta_{\mathcal{S}}^2k_{\mathcal{S}}\lambda^2}{m_{\mathcal{S}}^6}+\frac{16\eta_{\mathcal{S}}^4k_{\mathcal{S}}\lambda}{m_{\mathcal{S}}^8}$}&
        $\mathcal{O}_{H^4\mathcal{D}^4,3}^{(8)}$&
        \multicolumn{2}{c||}{$\frac{2\eta_{\mathcal{S}}^2}{m_{\mathcal{S}}^6}$}
        \\
        \cline{4-6}
        
        &
        \multicolumn{2}{c||}{$-\frac{\eta_{\mathcal{S}}^4\lambda_{\mathcal{S}}}{24m_{\mathcal{S}}^8}+\frac{8\eta_{\mathcal{S}}^6k_{\mathcal{S}}}{m_{\mathcal{S}}^{10}}$}&
        $\mathcal{O}_{\psi^2H^5}^{(8)}$&
        \multicolumn{2}{c||}{$-\frac{2\eta_{\mathcal{S}}^2k_{\mathcal{S}}Y_{\text{SM}}(1-2\lambda)}{m_{\mathcal{S}}^6}+\frac{4\eta_{\mathcal{S}}^4k_{\mathcal{S}}Y_{\text{SM}}}{m_{\mathcal{S}}^8}$}
        \\
	\hline
				
	$\mathcal{O}_{\psi^2H^3\mathcal{D}^2,1}^{(8)}$&
	$ -\frac{2\eta_{\mathcal{S}}^2k_{\mathcal{S}}Y_{\text{SM}}}{m_{\mathcal{S}}^6} $ &
        $\mathcal{O}_{\psi^4H^2,2}^{(8)}$ &
    \multicolumn{1}{c||}{$\frac{\eta_{\mathcal{S}}^2 k_{\mathcal{S}}Y_{\text{SM}}^2}{2m_{\mathcal{S}}^6}$}&
    
				$\mathcal{O}_{\psi^4H^2,1}^{(8)}$&
				$\frac{\eta_{\mathcal{S}}^2k_{\mathcal{S}}Y_{\text{SM}}^2}{m_{\mathcal{S}}^6}$

				\\

				\hline
				
			\end{tabular}
			\caption{\small Non-redundant SMEFT dimension-eight operators and their corresponding coefficients after integrating out the \textbf{Real Singlet Scalar} of Eq.~\eqref{eq:real_singlet_Lagrangian}.}
			\label{tab:rss-d8}
\end{table}
We use Eq.~\eqref{eq:redundant_to_nonredundant} to remove the redundancies from the above equation and rewrite them in the complete basis of Table~\ref{Tab:effective_ops_to_dim8}; the coefficients of non-redundant SMEFT dimension-eight operators are shown in Table~\ref{tab:rss-d8}. 
Expressed in the categories detailed above we arrive at
\begin{itemize}
    \item \underline{Tree-level contribution}: $\mathcal{O}_{H^6\mathcal{D}^2,1}^{(8)},\hspace{2mm} \mathcal{O}_{H}^{(8)},\hspace{2mm}\mathcal{O}_{\psi^2H^3\mathcal{D}^2,1}^{(8)}, \hspace{2mm}\mathcal{O}_{\psi^4 H^2,1}^{(8)},\hspace{2mm}\mathcal{O}_{\psi^2H^5}^{(8)}, \hspace{2mm}\mathcal{O}_{\psi^4H^2,2}^{(8)}$. The WCs corresponding to these operators are listed in Table~\ref{Tab:complex-quartet_dimension_8_coefficients}.
    \item \underline{Loop-induced and/or higher order contribution}: There is no redundancy at the dimen\-sion-six level, and no loop induced operators can be generated by the equation of motion. We can generate dimension-eight operators at one-loop-level itself by integrating out the heavy degree of freedom. This is beyond the scope of this paper, and we will leave this for future work. 
\end{itemize}

\subsection{Scalar Leptoquark}
Next, we consider the BSM model where the SM is extended by a scalar leptoquark, having quantum numbers $(3,2,1/6)$ under the SM gauge group $SU(3)_\text{C}\times SU(2)_\text{L} \times U(1)_\text{Y}$. This scenario has recently received lots of  attention as it can potentially address observed anomalies in $B$-meson decays~\cite{Lee:2021jdr,Sahoo:2015wya}. This model has also been analysed within the EFT framework, see Refs.~\cite{DasBakshi:2020pbf,Anisha:2020ggj}. We focus on the scalar interaction part of the Lagrangian for our discussion, which reads
\begin{multline} \label{eq:slq-lag}
       \mathcal{L}_\Theta  \supset  (\mathcal{D}_\mu \Theta^\dagger) (\mathcal{D}^\mu\Theta) - M_\Theta^2 (\Theta^\dagger\Theta)- \eta_{\Theta_1} (\Theta^\dagger\Theta) (H^\dagger H) -\eta_{\Theta_2} (\Theta^\dagger \tau^I \Theta) (H^\dagger \tau^I H) \\
       -\lambda_{\Theta_1} (\Theta^\dagger\Theta)^2
       -\lambda_{\Theta_2} (\Theta^\dagger \tau^I \Theta) (\Theta^\dagger \tau^I \Theta).
\end{multline}
No linear coupling of the Higgs field is present, and we do not obtain effective operators at tree-level. Note that although dimension-eight one-loop-level operators are beyond the scope of this work, we can still capture contributions to dimension-eight operators using our formalism. Table~\ref{Tab:leptoquark_SILH_coefficients} shows the one-loop generated operators and the corresponding WCs.
\begin{table}[!t]
\centering
\renewcommand{\arraystretch}{2.4} \scriptsize
\begin{tabular}{||c|c||c|c||c|c|}
	\hline
	\textsf{\quad SILH Op.} &
	\textsf{\quad Wilson Coef.} &
        \textsf{\quad SILH Op.} &
	\textsf{\quad Wilson Coef.} &
	\textsf{\quad SILH Op.} &
	\textsf{\quad Wilson Coef.} \\
	\hline
		
	$\mathcal{O}_{6}$ &
        $\biggl\{-\frac{\eta_{\Theta_1}^3}{16 \pi ^2 M_{\Theta}^2}-\frac{3 \eta_{\Theta_1}
        \eta_{\Theta2}^2}{256 \pi^2 M_{\Theta}^2}\biggr\}$ &
        $\mathcal{O}_{H}$ &
        $\biggl\{\frac{\eta_{\Theta_1}^2}{16 \pi^2 M_\Theta^2}\biggr\}$ &
        $\mathcal{O}_R$ &
        $\biggl\{\frac{\eta_{\Theta_2}^2}{128 \pi^2 M_\Theta^2}\biggr\}$ \\
	\hline
	
        $\mathcal{O}_T$ &
        $\biggl\{\frac{\eta_{\Theta_2}^2}{256 \pi^2 M_\Theta^2}\biggr\}$ &
        $ \mathcal{O}_{WW}$ &
        $\biggl\{\frac{\eta_{\Theta_1}}{128 \pi ^2 M_{\Theta}^2}\biggr\}$ &
        $\mathcal{O}_{BB}$ &
        $\biggl\{\frac{\eta_{\Theta_1}}{1152 \pi^2 M_{\Theta}^2}\biggr\}$ \\
        \hline

        $\mathcal{O}_{WB}$ &
        $\biggl\{\frac{\eta_{\Theta_2}}{768 \pi^2 M_\Theta^2}\biggr\}$ &
        $\mathcal{O}_{2W}$ &
        $\biggl\{\frac{g_W^2}{320 \pi^2 M_\Theta^2}\biggr\}$ &
        $\mathcal{O}_{2B}$&
        $\biggl\{\frac{g_Y^2}{2880 \pi^2 M_\Theta^2}\biggr\}$ \\
        \hline
        
	\end{tabular}
	\caption{\small WCs of dimension-six SMEFT operators in the SILH set for the {\bf{Scalar Leptoquark}} of Eq.~\eqref{eq:slq-lag}. The terms within braces $(\{\})$ denote the contribution from pure heavy loops.}
	\label{Tab:leptoquark_SILH_coefficients}
\end{table}
Table~\ref{Tab:leptoquark_dimension_8_coefficients_1} contains the WCs contribution coming from the loop induced operators at dimension-six only. The two operators quoted there is not an exhaustive list but serves the purpose of demonstrating that we can still obtain non-zero WCs from dimension-six operators without dimension-eight one-loop-level matching. Categorising the WCs as above we find
\begin{itemize}
    \item \underline{Tree-level contribution}: No tree-level contribution.
    \item \underline{Loop-induced and/or higher order contribution}: There is only loop induced contribution in this case. Table~\ref{Tab:leptoquark_dimension_8_coefficients_1} captures only a subset of operators which gets contribution from the lower dimension operators. The others are : $\mathcal{O}_{H^6\mathcal{D}^2,2}^{(8)},\hspace{2mm} \mathcal{O}_{\psi^2H^5}^{(8)},\hspace{2mm}\mathcal{O}_{\psi^2H^3\mathcal{D}^2,1}^{(8)},\\\hspace{2mm}\mathcal{O}_{\psi^2H^3\mathcal{D}^2,2}^{(8)}, 
    \hspace{2mm}\mathcal{O}_{\psi^4 H^2,1}^{(8)}, \hspace{2mm}\mathcal{O}_{\psi^4H^2,2}^{(8)},\,\mathcal{O}_{\psi^2H^4\mathcal{D},1}^{(8)},\, \mathcal{O}_{\psi^2H^4\mathcal{D},2}^{(8)},\,\mathcal{O}_{\psi^4\mathcal{D}H,1}^{(8)},\,\mathcal{O}_{\psi^4\mathcal{D}H,2}^{(8)}\,\mathcal{O}_{H^4 X^2}^{(8)},\\\,\mathcal{O}_{\psi^2X^2H}^{(8)}$. The full contribution can be accessed from the \mathematicanb{}.
\end{itemize}
    \begin{table}[!t]
    \centering\scriptsize
    \renewcommand{\arraystretch}{2.1}
        \begin{tabular}{|c|c|}
	\hline
	{\textsf{Operator}} &
	{\textsf{Wilson coefficients}} \\
	\hline

	$\mathcal{O}_{H^6\mathcal{D}^2,1}^{(8)}$&
	$\frac{g_W^4 g_Y^4 \lambda ^2}{14745600 \pi ^4 m_\Theta^4
        Y_{\text{SM}}^2}+\frac{g_W^4 g_Y^4 \lambda }{7372800 \pi ^4 m_\Theta^4
        Y_{\text{SM}}}+\frac{g_W^4 g_Y^4}{3686400 \pi ^4 m_\Theta^4}-\frac{\eta_{\Theta1}^2 g_W^4}{20480 \pi^4 m_{\Theta1}^4}+\frac{\eta_{\Theta2}^2
        g_W^4}{131072 \pi ^4 m_\Theta^4}$\\
   
        &$-\frac{\eta_{\Theta1}^2 g_W^4 \lambda}{20480 \pi ^4 m_\Theta^4 Y_{\text{SM}}}+\frac{\eta_{\Theta2}^2 g_W^4 \lambda }{327680 \pi ^4 m_{\Theta}^4 Y_{\text{SM}}}+\frac{g_W^8 \lambda ^2}{26214400
        \pi ^4 m_{\Theta}^4 Y_{\text{SM}}^2}-\frac{g_W^8 \lambda }{6553600 \pi ^4
        m_\Theta^4 Y_{\text{SM}}}-\frac{g_W^8}{5242880 \pi ^4 m_\Theta^4}$\\
   
        &$-\frac{\eta_{\Theta2}^2 g_Y^4}{184320 \pi ^4 m_\Theta^4}-\frac{\eta_{\Theta1}^2 g_Y^4 \lambda }{23040 \pi ^4 m_\Theta^4
        Y_{\text{SM}}}+\frac{\eta_{\Theta2}^2 g_Y^4 \lambda }{368640 \pi ^4 m_\Theta^4 Y_{\text{SM}}}+\frac{g_Y^8 \lambda ^2}{33177600 \pi ^4 m_\Theta^4
        Y_{\text{SM}}^2}+\frac{g_Y^8 \lambda }{4147200 \pi ^4 m_\Theta^4
        Y_{\text{SM}}}$ \\
   
        &$+\frac{\eta_{\Theta1}^2 \eta_{\Theta2}^2}{1024 \pi ^4 m_{\Theta}^4}-\frac{5 \eta_{\Theta2}^4}{65536 \pi ^4 m_{\Theta}^4}$\\
	  \hline
				
	\multirow{5}{*}{$\mathcal{O}_{H}^{(8)}$}&
	$-\frac{g_W^4 g_Y^4 \lambda }{1843200 \pi^4 m_\Theta^4}-\frac{g_W^4
        g_Y^4 \lambda^2}{1228800 \pi^4 m_\Theta^4 Y_{\text{SM}}}-\frac{11 g_W^4
        g_Y^4 \lambda^3}{7372800 \pi^4 m_\Theta^4 Y_{\text{SM}}^2}+\frac{\eta_\Theta^2 g_W^4 \lambda}{10240 \pi^4 m_\Theta^4}-\frac{3 \eta_{\Theta1}^3 g_W^4}{40960 \pi^4 m_\Theta^4}$ \\
   
        &$-\frac{9 \eta_{\Theta1} \eta_{\Theta2}^2 g_W^4}{655360 \pi^4 m_{\Theta}^4}+\frac{\eta_{\Theta2}^2
        g_W^4 \lambda }{65536 \pi^4 m_\Theta^4}-\frac{g_W^8 \lambda}{2621440 \pi^4
        m_\Theta^4}+\frac{\eta_{\Theta1}^2 g_W^4 \lambda^2}{10240 \pi^4
        m_\Theta^4 Y_{\text{SM}}}-\frac{3 \eta_{\Theta1}^3 g_W^4 \lambda }{40960 \pi^4 m_\Theta^4 Y_{\text{SM}}}$ \\
   
        &
        $-\frac{9 \eta_{\Theta1} \eta_{\Theta2}^2
        g_W^4 \lambda }{655360 \pi^4 m_\Theta^4 Y_{\text{SM}}}+\frac{7 \eta_{\Theta2}^2 g_W^4 \lambda^2}{163840 \pi^4 m_\Theta^4 Y_{\text{SM}}}-\frac{7 g_W^8\lambda ^2}{3276800 \pi ^4 m_\Theta^4 Y_{\text{SM}}}-\frac{23 g_W^8 \lambda^3}{13107200 \pi ^4 m_\Theta^4 Y_{\text{SM}}^2}+\frac{\eta_{\Theta2}^2 g_Y^4\lambda }{92160 \pi ^4 m_{\Theta}^4}$ \\
   
        &
        $+\frac{\eta_{\Theta1}^2 g_Y^4 \lambda^2}{11520 \pi ^4 m_\Theta^4 Y_{\text{SM}}}-\frac{\eta_{\Theta1}^3 g_Y^4\lambda }{15360 \pi ^4 m_{\Theta}^4 Y_{\text{SM}}}-\frac{\eta_{\Theta1} \eta_{\Theta2}^2 g_Y^4 \lambda }{81920 \pi ^4 m_\Theta^4 Y_{\text{SM}}}+\frac{\eta_{\Theta2}^2 g_Y^4 \lambda^2}{184320 \pi ^4 m_{\Theta}^4 Y_{\text{SM}}}-\frac{g_Y^8 \lambda ^2}{2073600 \pi ^4 m_{\Theta1}^4Y_{\text{SM}}}$\\
				
	&
	$+\frac{g_Y^8 \lambda ^3}{16588800 \pi ^4 m_{\Theta1}^4 Y_{\text{SM}}^2}-\frac{\eta_{\Theta1}^2            \eta_{\Theta2}^2 \lambda }{512 \pi ^4 m_{\Theta}^4}+\frac{3 \eta_{\Theta1}^3 \eta_{\Theta2}^2}{2048\pi^4m_{\Theta1}^4}+\frac{9 \eta_{\Theta1} \eta_{\Theta2}^4}{32768 \pi^4 m_{\Theta}^4}-\frac{5 \eta_{\Theta2}^4 \lambda }{32768 \pi ^4 m_\Theta^4}$ \\
	\hline
	
        \end{tabular}
	\caption{\small Contribution to dimension-eight operators from dimension-six operators. This table does not capture the full contribution. Full results are available from our \mathematicanb.}
	\label{Tab:leptoquark_dimension_8_coefficients_1}
    \end{table}
\section{Impact of dimension-eight operators on BSM scenarios}
\label{sec:classification}
Given the plethora of data available after LHC Run-II and Run-III, we broadly classify the above UV theories by investigating their low-energy phenomenology in this section, emphasising the relevance of the dimension-eight operators.

\begin{figure}
    \centering
    \includegraphics[scale=0.25]{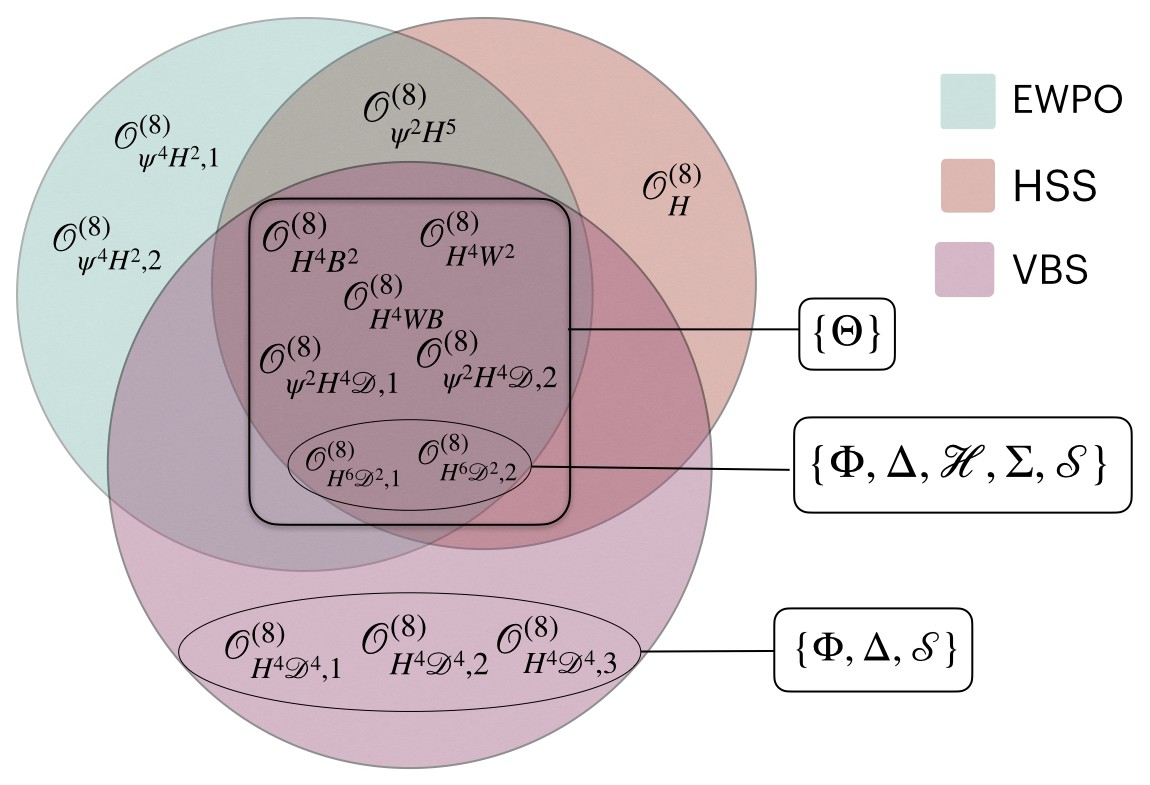}
    \caption{Interplay of different observables for the categorisation of complete models based on their sensitivity towards specific observable(s). \{$\Phi$,\,$\Delta$,\,$\mathcal{H}$,\,$\Sigma$,\,$\mathcal{S}$\} produce any one or both from the set \{$\mathcal{O}_{H^6\mathcal{D}^2,1}^{(8)}$,\,$\mathcal{O}_{H^6\mathcal{D}^2,2}^{(8)}$\} at tree-level. They contribute to all observables and are therefore severely constrained by EWPO. 
    \{$\mathcal{O}_{H^4\mathcal{D}^4}^{(1)},\,\mathcal{O}_{H^4\mathcal{D}^4}^{(2)},\,\mathcal{O}_{H^4\mathcal{D}^4}^{(3)}$\} are mainly constrained with VBS data, and filter out \{$\mathcal{H},\Sigma$\} which do not produce these operators.
    We highlight the operators produced by \{$\Theta$\} that contribute to all observables with a box, these are loop-suppressed. }
    \label{fig:chart_observables}
\end{figure}

Different observables and precision measurements provide strong discriminators between UV scenarios when using matched EFT results. In this sense, the dimension-eight effects provide quantitatively crucial additional information.
To gain a qualitative understanding of UV discrimination employing the results above, we consider three categories of experimental observables for guidance: 
(i) electroweak precision observables (EWPO), (ii) Higgs signal strength (HSS) measurements, and (iii) vector boson scattering (VBS) measurements. We analyse
the cases discussed in Sec.~\ref{sec:example}, reviewing the interplay of (i)-(iii) as shown in Fig.~\ref{fig:chart_observables}. 

The  characteristics of different models can be analysed by adjudging their responses to the following questions. First, one needs to note which effective operators  emerge from  each model under consideration. As the observables can be parametrized in terms of the effective operators, the observable-model correspondence can be set up directly. Based on that, one can classify  different UV models by carefully scrutinising the overlapping sets of operators contributing to a set of observables. The second question relates to the order of the perturbative expansion at which the operators are being produced. That will give a hint of their possible sensitivity towards the observables. Keeping all these points in mind, we have clubbed those models that show  degenerate sensitivity and prepared the different classes. Each class contains degenerate models with respect to their response to that particular observable in consideration.

The first question points to the need for new measurements with distinct features to correctly classify a wide range of complete models. The second one emphasises the ability of existing measurements to constrain the underlying UV parameter spaces determines the measurements' BSM UV sensitivity.

Such an observable-based categorisation has been studied recently (see e.g.~\cite{Bakshi:2020eyg}) 
for dimension-six operators up to one loop, mainly in the Warsaw basis. Bringing dimension-eight operators into the picture helps resolve model degeneracies at the dimension-six level even without introducing new measurements. However, since the number of independent structures increases rapidly beyond dimension six, we confine our discussion only to the structures emerging for the six scalar extensions discussed in Sec.~\ref{sec:x-validation} and Sec.~\ref{sec:example}.                                                                 
\subsection{Electroweak precision observables}
\label{subsec:EWPO}
The precise measurements of the electroweak observables naturally calls for improvements on the theoretical side. This can be achieved by performing theoretical computations at next-to-leading order (NLO) and by extending the effective series expansion. As described in Refs.~\cite{Dawson:2019clf,Dawson:2022bxd,Corbett:2021eux}, we organise the operators below into lists based on how they contribute to EWPO:
\begin{eqnarray}
\text{Dimension-six LO} &:& \{\mathcal{O}_{H\mathcal{D}},\,\mathcal{O}_{HWB},\,\mathcal{O}_{He},\,\mathcal{O}_{Hu},\,\mathcal{O}_{Hd},\,\mathcal{O}_{Hq}^{(1)},\,\mathcal{O}_{Hq}^{(3)},\,\mathcal{O}_{Hl}^{(1)},\nonumber\\
&&\,\mathcal{O}_{Hl}^{(3)},\,\mathcal{O}_{ll}\};\label{eq:dim-six-EWPO-LO}\\
\text{Dimension-six NLO} &:& \{\mathcal{O}_{\Box},\,\mathcal{O}_{HB},\,\mathcal{O}_{HW},\,\mathcal{O}_{W},\,\mathcal{O}_{uB},\,\mathcal{O}_{uW},\,\mathcal{O}_{ed},\,\mathcal{O}_{ee},\,\mathcal{O}_{eu},\nonumber\\
&&\,\mathcal{O}_{lu},\,\mathcal{O}_{ld},\,\mathcal{O}_{le},\,\mathcal{O}_{lq}^{(1)},\,\mathcal{O}_{lq}^{(3)},\,\mathcal{O}_{qe},\,\mathcal{O}_{qd}^{(1)},\,\mathcal{O}_{qq}^{(3)},\,\mathcal{O}_{qq}^{(1)},\,\mathcal{O}_{qu}^{(1)},\nonumber\\
&&\,\mathcal{O}_{ud}^{(1)},\,\mathcal{O}_{uu},\,\mathcal{O}_{dd}\}; \label{eq:dim-six-EWPO-NLO}\\
\text{Dimension-eight LO} &:& \{\mathcal{O}_{\psi^2H^5},\,\mathcal{O}_{H\mathcal{D},2}^{(1)},\,\mathcal{O}_{H\mathcal{D},2}^{(2)},\,\mathcal{O}_{\psi^2H^4\mathcal{D}}^{(1)},\,\mathcal{O}_{\psi^2H^4\mathcal{D}}^{(2)},\,\mathcal{O}_{H^4WB},\nonumber\\
&& \,\mathcal{O}_{\psi^4H^2}^{(1)},\,\mathcal{O}_{\psi^4H^2}^{(2)}\}.\label{eq:dim-eight-EWPO-LO}
\end{eqnarray}
    
Contrary to the $\Phi$ and $\Delta$ extensions, wherein \{$\mathcal{O}_{H\mathcal{D}}$\} and \{$\mathcal{O}_{H\mathcal{D}},\,\mathcal{O}_{ll}$\}, respectively, are produced at tree-level (see e.g. Refs.~\cite{Bakshi:2020eyg,Anisha:2021hgc}) and therefore provide the dominant contribution to the observable, other operators in Eq.~\eqref{eq:dim-six-EWPO-LO} are generally sourced by heavy one-loop insertions. 
Their subleading contributions are comparable to the ones resulting from the operators produced at the tree-level and contribute to the observable at NLO (noted in Eq.~\eqref{eq:dim-six-EWPO-NLO}). This implies that the BSM parameter space of such an extension is less sensitive to EWPO than other observables when only dimension-six is considered, which includes the models:\,\{$\mathcal{H},\,\Sigma,\,\mathcal{S}$\}. The situation can be remedied by taking dimension-eight operators shown in Eq.~\eqref{eq:dim-eight-EWPO-LO} into account. In this case, if these operators appear at the tree level, the constraints can be improved significantly. Hence based on sensitivity towards EWPO, we can divide all the models into two broad categories:
\begin{eqnarray}
\text{Class-A} &:& \{\Phi,\Delta,\mathcal{H},\Sigma, \mathcal{S}\},\nonumber\\
\text{Class-B} &:& \{\Theta\}\nonumber.
\end{eqnarray}

\subsection{Higgs signal strength measurements}
\label{subsec:HSS}
Higgs signal strengths (HSSs) are inherently connected to the interplay of fundamental mass generation in the SM and electroweak symmetry breaking. Therefore,
the analysis of the HSSs is a relevant discriminator in the space of Higgs sector extensions.
Certain dimension-eight operators imply non-negligible effects when constraining the BSM parameter space through HSS measurements, for instance, when the new physics occurs at a relatively low scale or if new couplings occur at tree-level after BSM states have been integrated out. According to the Refs.~\cite{Ellis:2018jhep,Anisha:2021hgc,Hays:2018zze}, the operators that affect the HSS measurements are listed below:
\begin{eqnarray}
\text{Dimension-six} &:& \{\mathcal{O}_H,\,\mathcal{O}_{H\Box},\,\mathcal{O}_{H\mathcal{D}},\,\mathcal{O}_{HB},\,\mathcal{O}_{HW},\,\mathcal{O}_{HWB},\,\mathcal{O}_{eH},\,\mathcal{O}_{uH},\nonumber\\
&&\,\mathcal{O}_{dH},\,\mathcal{O}_{He},\,\mathcal{O}_{Hu},\,\mathcal{O}_{Hd},\,\mathcal{O}_{Hq}^{(1)},\,\mathcal{O}_{Hq}^{(3)},\,\mathcal{O}_{Hl}^{(1)},\,\mathcal{O}_{Hl}^{(3)}\};\label{eq:dim-six-HSS}\\
\text{Dimension-eight} &:& \{\mathcal{O}_{H}^{(8)},\,\mathcal{O}_{H^6\mathcal{D}^2,1}^{(8)},\,\mathcal{O}_{H^6\mathcal{D}^2,2}^{(8)},\,\mathcal{O}_{\psi^2H^5},\,\mathcal{O}_{H^4B^2},\,\mathcal{O}_{H^4W^2},\,\mathcal{O}_{H^4WB}\}\label{eq:dim-eight-HSS}.
\end{eqnarray}

 Since the models \{$\Phi,\Delta,\mathcal{H},\mathcal{S}$\} produce subsets of these operators \{$\mathcal{O}_H,\,\mathcal{O}_{H\Box},\,\mathcal{O}_{H\mathcal{D}},\allowbreak\,\,\mathcal{O}_{uH},\,\mathcal{O}_{dH},\,\,\mathcal{O}_{eH}$\} at tree-level, while for \{$\Sigma, \Theta$\} they are generated at one-loop (see Ref.~\cite{Bakshi:2020eyg}), these models are seemingly less sensitive to HSS measurements at dimension-six. Following a similar approach as the one described in Sec.~\ref{subsec:EWPO}, we can infer that the impact of the operators given in Eq.~\eqref{eq:dim-eight-HSS} should be considered to properly explore the parameter space of \{$\Sigma, \Theta$\}. The fact that $\Sigma$ generates $\mathcal{O}_{H}^{(8)}$, $\mathcal{O}_{H^6\mathcal{D}^2,1}^{(8)}$ and $\mathcal{O}_{H^6\mathcal{D}^2,2}^{(8)}$ at tree-level, as shown in Table~\ref{Tab:complex-quartet_dimension_8_coefficients}, 
 leads to similar a priori sensitivity of Higgs measurements as for
 \{$\Phi,\Delta,\mathcal{H},\mathcal{S}$\}. HSSs therefore discriminate:
\begin{eqnarray}
\text{Class-A} &:& \{\Phi,\Delta,\mathcal{H},\Sigma, \mathcal{S}\},\nonumber\\
\text{Class-B} &:& \{\Theta\}\nonumber.
\end{eqnarray}

\subsection{Vector boson scattering measurements}
\label{subsec:vbs}
Vector boson scattering measurements have been very crucial in the study of the electroweak sector, particularly in constraining  anomalous gauge couplings, which have been discussed in detail in
SMEFT at dimension-six~\cite{Gomez:2019epj,Araz:2021jhep,Banerjee:2019twi,Dedes:2021prd}. Individual bounds on dimension-eight couplings have been derived from VBS data as well~\cite{Lang:2021hnd,CMS:2020meo,Guo:2019agy}. At dimension-six, a total of nine operators contribute to the modification of observables, through gauge-self couplings (\{$\mathcal{O}_{W}$\}), gauge-Higgs couplings (\{$\mathcal{O}_{H\mathcal{D}},\mathcal{O}_{HW},\mathcal{O}_{HB},\mathcal{O}_{HWB}$\}) and fermion-gauge couplings (\{$\mathcal{O}_{Hl}^{(1)},\mathcal{O}_{Hl}^{(3)},\mathcal{O}_{Hq}^{(1)},\mathcal{O}_{Hq}^{(3)}$\}). Among these, $\mathcal{O}_{H\mathcal{D}}$ and the two-fermionic operators are mainly constrained from EWPO observables~\cite{Ellis:2018gqa}, while the operators  \{$\mathcal{O}_{W},\mathcal{O}_{HW},\mathcal{O}_{HB},\mathcal{O}_{HWB}$\} remain to be constrained by VBS measurements~\cite{Ethier:2021epj}. These are typically produced at one-loop level (see Ref.~\cite{Bakshi:2020eyg}). For a low enough cut-off scale, these can be comparable to dimension-eight tree-level contributions
\begin{eqnarray}
\text{Dimension-six} &:& \{\mathcal{O}_{W},\,\mathcal{O}_{HB},\,\mathcal{O}_{HW},\,\mathcal{O}_{HWB}\};\\
\text{Dimension-eight} &:& \{\mathcal{O}_{H\mathcal{D},2}^{(1)},\,\mathcal{O}_{H\mathcal{D},2}^{(2)},\,\mathcal{O}_{H^4\mathcal{D}^4}^{(1)},\,\mathcal{O}_{H^4\mathcal{D}^4}^{(2)},\,\mathcal{O}_{H^4\mathcal{D}^4}^{(3)}\}.
\end{eqnarray}
 We note that, $\{\mathcal{O}_{H\mathcal{D},2}^{(1)},\,\mathcal{O}_{H\mathcal{D},2}^{(2)}\}$ contribute to the EW sector through the modification of gauge boson masses (see Appendix D of Ref.~\cite{Hays:2018zze} for more details). Thus they are mostly constrained by EWPO. Models that produce the full or a subset of the rest of the mentioned dimension-eight operators (i.e., \{$\Phi$,$\Delta$,$\mathcal{S}$\}), can be efficiently constrained by VBS measurements: 
\begin{eqnarray}
\text{Class-A} &:& \{\Phi,\Delta,\mathcal{S}\},\nonumber\\
\text{Class-B} &:& \{\mathcal{H},\Sigma,\Theta\}.
\end{eqnarray}		

\section{Conclusions}
\label{sec:conc}
The indirect search for new physics using EFT, while providing an ingenious way to uncover the physics that might lie just beyond our reach, faces several critical challenges when tracing constraints to possible complete and renormalisable UV scenarios. Moreover, since one encounters new signatures at dimension eight that may unravel the microscopic nature of new interactions, including their effects can become a vital question when looking for new physics in a model-independent way. 
However, performing a global analysis of the entire parameter space of dimensions six and eight SMEFT is unrealistic. Broad model-dependent correlations can then help to hone the sensitivity to new interactions. This requires a transparent and effective way to perform matching to new physics scenarios beyond dimension six. In this work, we have explored these two issues in detail.

We present an easy-to-implement approach to compute the dimension-eight matching coefficients, capturing loop effects consistently. Furthermore, the method employs EOMs instead of the traditional field-redefinition formalism.
The ``missing piece" of the EOM that elevates it to a similar footing with field redefinition is included in a model-independent manner. It must be stressed that, while removing the redundant structures at dimension six, we obtain one-loop or two-loop equivalent contributions to dimension-eight structures due to the interference among the pieces generated at the tree and one-loop order. 

We have applied this method to six different scalar extensions of the SM at one-loop order of the dimension-six coefficients considering both heavy-heavy and heavy-light loop propagators, validating our approach against results documented in the literature.    
Finally, we have clarified the relevance of dimension-eight operators for classifying UV-complete models given Higgs and electroweak measurements and VBS data.

\section*{Acknowledgements}
The authors thank Wrishik Naskar for comments after carefully reading this work during its draft stage. We also acknowledge helpful comments from Dave Sutherland. The work of U.B., J.C., and S.U.R. is supported by the Science and Engineering Research Board, Government of India, under the agreement SERB/PHY/2019501 (MATRICS). C.E. is supported by the STFC under grant ST/T000945/1, by the Leverhulme Trust under grant RPG-2021-031, and the IPPP Associateship Scheme. M.S. is supported by the STFC under grant ST/P001246/1.

\appendix

\begin{table}[!b]
	\centering \scriptsize
	\renewcommand{\arraystretch}{2.2}
     \resizebox{\textwidth}{!}{\begin{tabular}{|c|c|c|c|}
		\hline
		\textsf{\quad Operator}$\quad$&
		\textsf{\quad Op. Structure}$\quad$&
		\textsf{\quad Operator}$\quad$&
		\textsf{\quad Op. Structure}$\quad$\\
		\hline
		
		$\mathcal{Q}_{H}$&
		$\frac{1}{2}\partial_{\mu}(H^{\dagger}H)\partial^\mu(H^{\dagger}H)$&
		$\mathcal{Q}_6$&
		$|H|^6$\\
		\hline

		$\mathcal{Q}_R$&
        $(H^{\dagger}H)(\mathcal{D}_{\mu}H^{\dagger}D^{\mu}H)$&
        $\mathcal{Q}_T$&
		$\frac{1}{2}\big[H^{\dagger}\overleftrightarrow{\mathcal{D}}^{\mu}H\big]\big[H^{\dagger}\overleftrightarrow{\mathcal{D}}^{\mu}H\big]$\\
		\hline
		
		$\mathcal{Q}_{\mathcal{D}}$&
		$(\mathcal{D}^2H^{\dagger})(\mathcal{D}^2H)$&
		$\mathcal{Q}_{2W}$&
		$-\frac{1}{2} \big(\mathcal{D}_{\mu}W^I_{\mu\nu}\big)^2$\\
		\hline
		
		$\mathcal{Q}_{2B}$&
		$-\frac{1}{2} \big(\partial_{\mu}B_{\mu\nu}\big)^2$&
		$\mathcal{Q}_{W}$&
		$ig_{_W}(H^{\dagger}\tau^a\overleftrightarrow{\mathcal{D}}^{\mu}H)\mathcal{D}^{\nu}W^a_{\mu\nu}$\\
		\hline
		
		$\mathcal{Q}_{B}$&
		$ig_{_Y}(H^{\dagger}\overleftrightarrow{\mathcal{D}}^{\mu}H)\partial^{\nu}B_{\mu\nu}$&
		$\mathcal{Q}_{WW}$&
		$g_{_W}^2(H^{\dagger}H)W^a_{\mu\nu}W^{a,\mu\nu}$\\
		\hline
		
		$\mathcal{Q}_{BB}$&
		$g_{_Y}^{2}(H^{\dagger}H)B_{\mu\nu}B^{\mu\nu}$&
		$\mathcal{Q}_{WB}$&
		$2g_{_W}g_{_Y}(H^{\dagger}\tau^aH)W^a_{\mu\nu}B^{\mu\nu}$\\
		\hline

	\end{tabular}}
	\caption{\small Dimension-six operator structures in the SILH set.}
	\label{Tab:effective_ops_dim6}
\end{table}

\section{Relevant operator structures}\label{app:op-structures}	
Here we discuss the operator structure that differs from the \emph{Green's set} as defined in Ref.~\cite{Gherardi:2020det}. At dimension-six there are four structures in the $\Phi^4 D^2$ operator class. The operators are the following:
\begin{align}
    \mathcal{O}_{H\square}= (H^\dagger H) \square (H^\dagger H),\;\; & \mathcal{O}_{H\mathcal{D}} = |H^\dagger \mathcal{D_\mu}H|^2 \nonumber\\
    \mathbf{\mathcal{O}_{H\mathcal{D}}^{\prime}= (H^\dagger H)(\mathcal{D}^\mu H^\dagger \mathcal{D_\mu}H)},\;\; & \mathbf{\mathcal{O}_{H\mathcal{D}}^{\prime\prime}= (H^\dagger H) \mathcal{D}^\mu (H^\dagger \, i\overleftrightarrow{\mathcal{D}}_\mu H) }.
\end{align}
Among the above structures, the first two $(\mathcal{O}_{H\square},\, \mathcal{O}_{H\mathcal{D}} )$ are considered to be the independent and part of the complete \emph{Warsaw basis}. We can ignore the last one, $\mathcal{O}_{H\mathcal{D}}^{\prime\prime}$, which is CP-violating and does not appear in our analysis. The redundant operator $\mathcal{O}_{H\mathcal{D}}^{\prime}$ is the important structure for our analysis. In order to remove this redundancy, we derive the contribution to higher dimension i.e., dimension-eight in our case. Instead of using this exact structure, we use the following relation to convert it to a suitable form
\begin{align}
    \mathbf{(H^{\dagger}H)(\mathcal{D}_{\mu}H^{\dagger}D^{\mu}H)} = \frac{1}{2}\big[(H^{\dagger}H)\square(H^{\dagger}H)-\mathbf{(H^{\dagger}H)(\mathcal{D}^2 H^{\dagger}H+H^{\dagger}\mathcal{D}^2 H)}\big]
\end{align}
Since we are replacing one redundant structure with another that is related to the former by the integration by parts, we term it a \emph{Green's set}-like structure. We collect all relevant dimension-six operators in Table~\ref{Tab:effective_ops_dim6}. Dimension-eight can be found in Table~\ref{Tab:effective_ops_dim8}.

\begin{table}[!t]
	\centering \scriptsize
	\renewcommand{\arraystretch}{2.2}
     \resizebox{\textwidth}{!}{\begin{tabular}{|c|c|c|c|}
		\hline
		\textsf{\quad Operator}$\quad$&
		\textsf{\quad Op. Structure}$\quad$&
		\textsf{\quad Operator}$\quad$&
		\textsf{\quad Op. Structure}$\quad$\\
		\hline
		
		$\mathcal{O}_{H}^{(8)}$&
		$(H^{\dagger}H)^4$&
		$\mathcal{O}_{\psi^2H^5}^{(8)}$&
		$(H^{\dagger}H)(\overline{\psi}_i\psi_jH)$\\
		\hline

		$\mathcal{O}_{H^6\mathcal{D}^2,1}^{(8)}$&
        $(H^{\dagger}H)^2(\mathcal{D}_{\mu}H^{\dagger}\mathcal{D}^{\mu}H)$&
        $\mathcal{O}_{H^6\mathcal{D}^2,2}^{(8)}$&
		$(H^{\dagger}H)(H^{\dagger}\mathcal{D}_{\mu}H\,\mathcal{D}^{\mu}H^{\dagger}H)$\\
		\hline
		
		$\mathcal{O}_{\psi^2H^4\mathcal{D},1}^{(8)}$&
		$i(H^{\dagger}H)(\psi_i\gamma_\mu\psi_j)(H^{\dagger}\overleftrightarrow{\mathcal{D}}_{\mu}H)$&
		$\mathcal{O}_{\psi^2H^4\mathcal{D},2}^{(8)}$&
		$i(H^{\dagger}H)(\psi_i\tau^I\gamma_\mu\psi_j)(H^{\dagger}\overleftrightarrow{\mathcal{D}}^I_{\mu}H)$\\
		\hline
		
		$\mathcal{O}_{\psi^4\mathcal{D}H,1}^{(8)}$&
		$i(\overline{\psi}_i\gamma^{\mu}\psi_j)[(\overline{\psi}_k\psi_l)\mathcal{D}_{\mu}H]$&
		$\mathcal{O}_{\psi^4\mathcal{D}H,2}^{(8)}$&
		$i(\overline{\psi}_i\gamma^{\mu}\tau^I\psi_j)[(\overline{\psi}_k\psi_l)\tau^I\mathcal{D}_{\mu}H]$\\
		\hline
		
		$\mathcal{O}_{H^4B^2}^{(8)}$&
		$(H^{\dagger}H)^2B_{\mu\nu}B^{\mu\nu}$&
		$\mathcal{O}_{H^4W^2}^{(8)}$&
		$(H^{\dagger}H)^2W_{\mu\nu}^I W^{\mu\nu,I}$\\
		\hline
		
		$\mathcal{O}_{H^4WB}^{(8)}$&
		$(H^{\dagger}H)(H^{\dagger}\tau^I H)W_{\mu\nu}^I B^{\mu\nu}$&
		$\mathcal{O}_{\psi^2WBH}^{(8)}$&
		$(\overline{\psi}_i\psi_j)\tau^I H W_{\mu\nu}^I B^{\mu\nu}$\\
		\hline
		
		$\mathcal{O}_{\psi^2B^2H}^{(8)}$&
		$(\overline{\psi}_i\psi_j)HB_{\mu\nu}B^{\mu\nu}$&
		$\mathcal{O}_{\psi^2W^2H}^{(8)}$&
		$(\overline{\psi}_i\psi_j)H W_{\mu\nu}^I W^{\mu\nu,I}$\\
		\hline

		$\mathcal{O}_{\psi^2H^3\mathcal{D}^2,1}^{(8)}$&
		$(\overline{\psi}_i\psi_j H) (\mathcal{D}_{\mu}H^{\dagger}\mathcal{D}^{\mu}H)$&
		$\mathcal{O}_{\psi^2H^3\mathcal{D}^2,2}^{(8)}$&
		$(H^{\dagger}\mathcal{D}_{\mu}H)(\overline{\psi}_i\psi_j\mathcal{D}^{\mu}H)$\\
		\hline

		$\mathcal{O}_{\psi^4H^2,1}^{(8)}$&
		$(H^{\dagger}H)(\overline{\psi}_i\psi_j\overline{\psi}_k\psi_l)$&
		$\mathcal{O}_{\psi^4H^2,2}^{(8)}$&
		$(\overline{\psi}_i \psi_j H) (\widetilde{H}^{\dagger}\overline{\psi}_k\psi_l)$\\
		\hline

	\end{tabular}}
	\caption{\small Structures of the dimension-eight operators in the non-redundant basis discussed throughout this work. $\psi$ denotes any SM fermion $(\psi \in \{q,u,d,l,e\})$.}
	\label{Tab:effective_ops_dim8}
\end{table}

\section{Renormalisable SM Lagrangian and EOM}\label{app:SM-lag-eom}
The renormalisable SM Lagrangian is
\begin{eqnarray}\label{eq:sm-lag}
\mathcal{L}_{\text{SM}} &= -\cfrac{1}{4} G_{\mu \nu}^A G^{A\mu \nu}-\cfrac{1}{4} W_{\mu \nu}^I W^{I \mu \nu} -\cfrac{1}{4} B_{\mu \nu} B^{\mu \nu}
+ (\mathcal{D}_\mu H^\dagger)(\mathcal{D}^\mu H)
+\sum\limits_{\psi=q,u,d,l,e} \overline \psi\, i \slashed{\mathcal{D}} \, \psi\, \nonumber\\
&-\lambda \left(H^\dagger H\right)^2- \biggl[ Y_d\, H^{\dagger j} \overline d\, q_{j} +  Y_u\, \widetilde{H}^{\dagger j} \overline u\, q_{j} + Y_e\, H^{\dagger j} \overline e\,   l_{j} + \hbox{h.c.}\biggr].
\end{eqnarray}
Here, we have ignored the (negative) mass term for the Higgs field which is not relevant for our analysis. We can calculate the EOMs for various fields using Eq.~\eqref{eq:sm-lag}. These EOMs constitute the first-order approximation of the full EOM, and they can be used to transform one set to another at a given mass dimension. The first-order EOM for Higgs field is~\cite{Jenkins:2013zja,Barzinji:2018xvu}
\begin{align}
  & \mathcal{D}^2 H_k +2 \lambda (H^\dagger H) H_k + \mathcal{Y}_k =0 \nonumber\\
\text{where, } &\mathcal{Y}_k=Y_u^\dagger\, \overline{q}^j\, u \epsilon_{jk} + Y_d\, \overline{d}\, q_k +Y_e\, \overline{e}\, l_k\,,
\label{eom-H}
\end{align}
the EOMs for the fermions are~\cite{Jenkins:2013zja,Barzinji:2018xvu}
\begin{align}
i\slashed{\mathcal{D}}\, q_j &= Y_u^\dagger\, u\, \widetilde H_j + Y_d^\dagger\, d\, H_j \,, &
i\slashed{\mathcal{D}}\,  d &= Y_d\,  q_j\, H^{\dagger\, j} \,, &
i\slashed{\mathcal{D}}\, u &= Y_u\, q_j\, \widetilde H^{\dagger\, j}\,, \nn\\
i\slashed{\mathcal{D}} \, l_j &= Y_e^\dagger\, e  H_j  \,, &
i\slashed{\mathcal{D}} \, e &= Y_e\, l_j H^{\dagger\, j}\,,
\label{eompsi}
\end{align}
with Yukawa couplings $Y_i\equiv Y_{\text{SM},i}$.
The EOMs for the gauge fields are~\cite{Jenkins:2013zja,Barzinji:2018xvu}
\begin{align}
\label{eomX}
\left[\mathcal{D}^\alpha , G_{\alpha \beta} \right]^A &= g_G  \sum_{\psi=u,d,q} \overline \psi \, T^A \gamma_\beta  \psi,\nonumber \\
\left[\mathcal{D}^\alpha , W_{\alpha \beta} \right]^I &= g_W  \Bigg(\frac{1}{2} \overline{q} \, \tau^I \gamma_\beta  q + \frac{1}{2} \overline{l} \, \tau^I \gamma_\beta  l + \frac{1}{2} H^\dagger \, i\overleftrightarrow{\mathcal{D}}_\beta^I H \Bigg), \\
\left[\mathcal{D}^\alpha, B_{\alpha \beta}\right] &= g_Y \Bigg( \sum_{\psi=u,d,q,e,l} \overline{\psi} \, \mathrm{y}_i \gamma_\beta  \psi + H^\dagger \, i\overleftrightarrow{\mathcal{D}}_\beta H \Bigg),\nonumber
\end{align}
 where  $\mathrm{y}_i$ denotes the $U(1)_{\text{Y}}$ hypercharges of the fermions. We have also used the following the notation \cite{Grzadkowski:2010es,Jenkins:2013zja},
\begin{eqnarray}
H^\dagger \, i\overleftrightarrow{\mathcal{D}}_\beta H &=& i H^\dagger (\mathcal{D}_\beta H) - i (\mathcal{D}_\beta H^\dagger) H, \\ \nonumber
H^\dagger \, i\overleftrightarrow{\mathcal{D}}_\beta^I H &=& i H^\dagger \tau^I (\mathcal{D}_\beta H) - i (\mathcal{D}_\beta H^\dagger)\tau^I H\,,
\end{eqnarray}
to write the operators in the well-known compact forms.
\bibliographystyle{jhep}
\bibliography{ref.bib}
\end{document}